\pdfoutput=1
\documentclass[11pt,twoside,a4paper,cmspaper,final,collab]{cms-tdr}
\def\svnVersion{955628d}\def\svnDate{2026/04/30}\def\cmsCernNoTag{CERN-EP-2025-212}\def\cmsCernDate{\today}\def\cmsMessage{Published in the Journal of High Energy Physics as \href{http://dx.doi.org/10.1007/JHEP04(2026)133}{\doi{10.1007/JHEP04(2026)133}.}}
\begin{document}\cmsNoteHeader{EXO-23-004}

\newlength\cmsFigWidth
\ifthenelse{\boolean{cms@external}}{\setlength\cmsFigWidth{0.85\columnwidth}}{\setlength\cmsFigWidth{0.4\textwidth}}
\providecommand{\CL}{CL\xspace}
\newcommand{\mjj}{\ensuremath{m_{\mathrm{jj}}}\xspace}
\newcommand{\detajj}{\ensuremath{\abs{\Delta\eta}}\xspace}
\newcommand{\Qstar}{\ensuremath{\PQq^*}\xspace}
\newcommand{\RunLumi}{117\fbinv}
\newcommand{\jecUncert}{2\%\xspace}
\newcommand{\CALOminMjjCut}{0.5\TeV}
\newcommand{\mDM}{\ensuremath{M_{\text{DM}}}\xspace}
\newcommand{\gDM}{\ensuremath{g_{\mathrm{DM}}}\xspace}
\newcommand{\gq}{\ensuremath{g_{\mathrm{q}}}\xspace}
\newcommand{\gqPrime}{\ensuremath{g_{\mathrm{q}}'}\xspace}
\newcommand{\mMed}{\ensuremath{M_{\text{med}}}\xspace}
\providecommand{\HT}{\ensuremath{H\_\mathrm{T}}\xspace}
\newcommand{\pvalue}{\ensuremath{p\text{-value}}}
\newcommand{\pvalues}{\ensuremath{p\text{-values}}}

\cmsNoteHeader{EXO-23-004}

\title{Search for dijet resonances with data scouting in proton-proton collisions at \texorpdfstring{$\sqrt{s} = 13\TeV$}{sqrt(s) = 13 TeV}}

\date{\today}

\abstract{A search is presented for narrow resonances, with a mass between 0.6 and 1.8\TeV, decaying to pairs of jets, in proton-proton collisions at $\sqrt{s} = 13\TeV$. The search is performed using dijets that are reconstructed, selected, and recorded in a compact form by the high-level trigger in a technique referred to as ``data scouting", from data collected in 2016--2018 corresponding to an integrated luminosity of \RunLumi. The dijet mass spectra are well described by a smooth parameterization, and no significant evidence for the production of new particles is observed. Model-independent upper limits are presented on the product of the cross section, branching fraction, and acceptance for the individual cases of narrow quark-quark, quark-gluon, and gluon-gluon resonances, and are compared to the predictions from a variety of models of narrow dijet resonance production. The upper limit on the coupling of a dark matter mediator to quarks is presented as a function of the mediator mass. The sensitivity of this search goes beyond what is expected from statistical scaling with the integrated luminosity alone, as a consequence of the use of fewer parameters in the background function within a more robust statistical procedure.}

\hypersetup{%
pdfauthor={CMS Collaboration},%
pdftitle={Search for dijet resonances with data scouting in proton-proton collisions at \texorpdfstring{$\sqrt{s} = 13\TeV$}{sqrt(s) = 13 TeV}},%
pdfsubject={CMS},%
pdfkeywords={CMS, dijet}}

\maketitle 

\section {Introduction}

Many theories that extend the standard model (SM) of particle physics predict new particles that decay to pairs of partons, quarks (q) and gluons (g), that hadronize into jets, which manifest themselves as dijet resonances~\cite{ref_gauge,ref_axi,ref_qstar,ref_diquark,Baur:1989kv,ref_coloron,ref_rsg,Cullen:2000ef,Han:2010rf,Chivukula:2013xla,Chala:2015ama,Abdallah:2015ter,Abercrombie:2015wmb,BOVEIA2020100365}. These resonances would appear in experiments as an excess of dijet events on top of a smoothly falling background in the dijet mass (\mjj) spectrum. Model-independent searches for these resonances have been conducted at the CERN Large Hadron Collider (LHC) in the following intervals of the dijet mass: high (above 2\TeV), medium (between 0.5 and 2\TeV), and low (below 0.5\TeV). The high-mass searches~\cite{ATLAS2010,Khachatryan:2010jd,Aad:2011aj,Chatrchyan2011123,Aad201237,ATLAS:2012pu,CMS:2012yf,Chatrchyan:2013qhXX,Aad:2014aqa,Khachatryan:2015sja,ATLAS:2015nsi,Khachatryan:2015dcf,Sirunyan:2016iap,Aaboud:2017yvp,Sirunyan:2018xlo,Aad:2019hjw,Sirunyan:2019vgj} rely on events selected by jet triggers with high thresholds, and the individually resolved jets are reconstructed offline. The low-mass searches use events selected by triggering on initial-state radiation of photons or gluons~\cite{CMS:2019xai,ATLAS:2018hbc,CMS:2019emo}, and the highly Lorentz-boosted, merged dijets are also reconstructed offline. This analysis is one of the medium-mass searches~\cite{Sirunyan:2018xlo,Sirunyan:2016iap,Khachatryan:2016ecr,ATLAS:2018qto}, which relies on events that are reconstructed, selected, and recorded in a compact form by the high-level trigger (HLT) in a technique referred to as ``data scouting"~\cite{2024scoutingRun3}. The compact form and the faster online reconstruction of the individually resolved jets using only calorimeter information allow a higher bandwidth of the trigger and a lower threshold in dijet mass than the high-mass searches. The scouting data sets offer a unique way to probe previously unexplored regions in the dijet mass at lower resonance couplings to quarks and gluons, not otherwise accessible by the full data sets with offline reconstruction.

Dark matter (DM) mediators, hypothetical particles that transmit an interaction between quarks and DM particles, can appear as dijet resonances~\cite{Chala:2015ama,Abercrombie:2015wmb,Abdallah:2015ter,BOVEIA2020100365}. Since the interaction between DM and ordinary matter is unknown, many models exist. As a benchmark, we choose the simplified model~\cite{BOVEIA2020100365} recommended by the LHC DM Working Group to explore the sensitivity to DM from LHC searches. In this model, the DM particle is assumed to be a Dirac fermion, and the particle mediating the interaction is exchanged in the $s$-channel. This approach provides a coherent framework for analyzing potential interactions and setting limits on DM properties. The simplified model is characterized by four parameters: the DM particle mass ($\mDM$), the mediator mass ($\mMed$), the universal mediator coupling to quarks ($\gq$), and the mediator coupling to DM particles ($\gDM$). The considered mediator only couples to $\PQq\PAQq$ or DM particle pairs. If $\mDM>\mMed/2$, the on-shell mediator cannot decay to DM particles, and the dijet cross section ($\sigma$) in this case becomes identical to that in a leptophobic $\PZpr$ model, in which the universal coupling to quarks is $\gqPrime$.

This paper describes a model-independent search for narrow $s$-channel dijet resonances with masses between 0.6 and 1.8\TeV. The search is based on proton-proton ($\Pp\Pp$) collision data at $\sqrt{s} = 13\TeV$ collected with data-scouting triggers in 2016--2018 and corresponding to an integrated luminosity of \RunLumi. We estimate signal significances and set limits on the production cross sections of new particles decaying to the parton pairs $\PQq\PQq$ (or $\PQq\PAQq$), $\PQq\Pg$, and $\Pg\Pg$. We then use these limits to exclude models of new gauge bosons $\PWpr$ and $\PZpr$~\cite{ref_gauge}, axigluons~\cite{ref_axi}, excited quarks~\cite{ref_qstar,Baur:1989kv}, scalar diquarks~\cite{ref_diquark}, colorons~\cite{ref_coloron}, and color-octet scalars~\cite{Han:2010rf} over the complete mass range considered. In addition, the 95\% confidence level (\CL) upper limit on the coupling strength $\gq$ of DM mediators is presented as a function of $\mMed$. The tabulated results for this analysis are provided in the HEPData record~\cite{HEPData}.

\section{The CMS detector and the scouting triggers}
\label{sec:detectorAndTrigger}

A detailed description of the CMS detector and its coordinate system, including definitions of the azimuthal angle $\phi$ and pseudorapidity $\eta$, is given in Refs.~\cite{refCMS,Hayrapetyan_2024_Run3}. The central feature of the CMS apparatus is a superconducting solenoid of 6\unit{m} internal diameter providing an axial magnetic field of 3.8\unit{T}. Within the solenoid volume are located the silicon pixel and strip tracker ($\abs{\eta}<2.7$), and the barrel and endcap calorimeters ($\abs{\eta}<3.0$), where these latter detectors consist of a lead tungstate crystal electromagnetic calorimeter (ECAL), and a brass and scintillator hadron calorimeter. An iron and quartz-fiber hadron calorimeter is located in the forward region ($3.0<\abs{\eta}<5.0$), outside the solenoid volume. The muon detection system covers $\abs{\eta}<2.4$ with up to four layers of gas-ionization chambers installed outside the solenoid and interleaved with layers of the steel flux-return yoke.

In this search we utilize jets formed from energy deposits in the calorimeters, denoted as Calo-jets, in order to benefit from the lowest possible trigger thresholds with the scouting data set. To reconstruct online jets used by the trigger, we employ the anti-\kt algorithm~\cite{Cacciari:2005hq,Cacciari:2008gp} with a distance parameter of 0.4, as implemented in the \textsc{FastJet} package~\cite{Cacciari:2011ma}. An event-by-event correction based on the jet area~\cite{Cacciari:2007fd,Khachatryan:2016kdb} is applied to the jet energy to remove the estimated contribution from additional collisions in the same or adjacent bunch crossings (pileup).

Events are recorded using a two-tier trigger system~\cite{Khachatryan:2016bia,Hayrapetyan_2024_Run2}. Events satisfying loose jet requirements at the first level (L1) trigger are examined by the HLT. When an event satisfies the data scouting requirement, the Calo-jets reconstructed at the HLT are saved, along with the event energy density and the missing transverse momentum reconstructed from the calorimeter. The energy density is defined as the median calorimeter energy per unit area calculated for each event in a grid of $\eta$-$\phi$ cells~\cite{Khachatryan:2016kdb}. 

We apply triggers, which require \HT to exceed a threshold, where \HT is the scalar sum of transverse momentum (\pt) for all jets in the event with $\abs{\eta}<3.0$. The shorter time for event reconstruction of calorimeter quantities and the reduced event size recorded for these events allow a reduced \HT threshold compared to the searches not utilizing scouting data sets~\cite{2024scoutingRun3}. Calo-jets with $\pt>40\GeV$ are used to compute \HT, and the threshold is set to $\HT>250\GeV$. For comparison, when not using scouting, the lowest threshold trigger requires $\HT>1050\GeV$.

\begin{figure*}[!htb]
\includegraphics[width=0.32\textwidth]{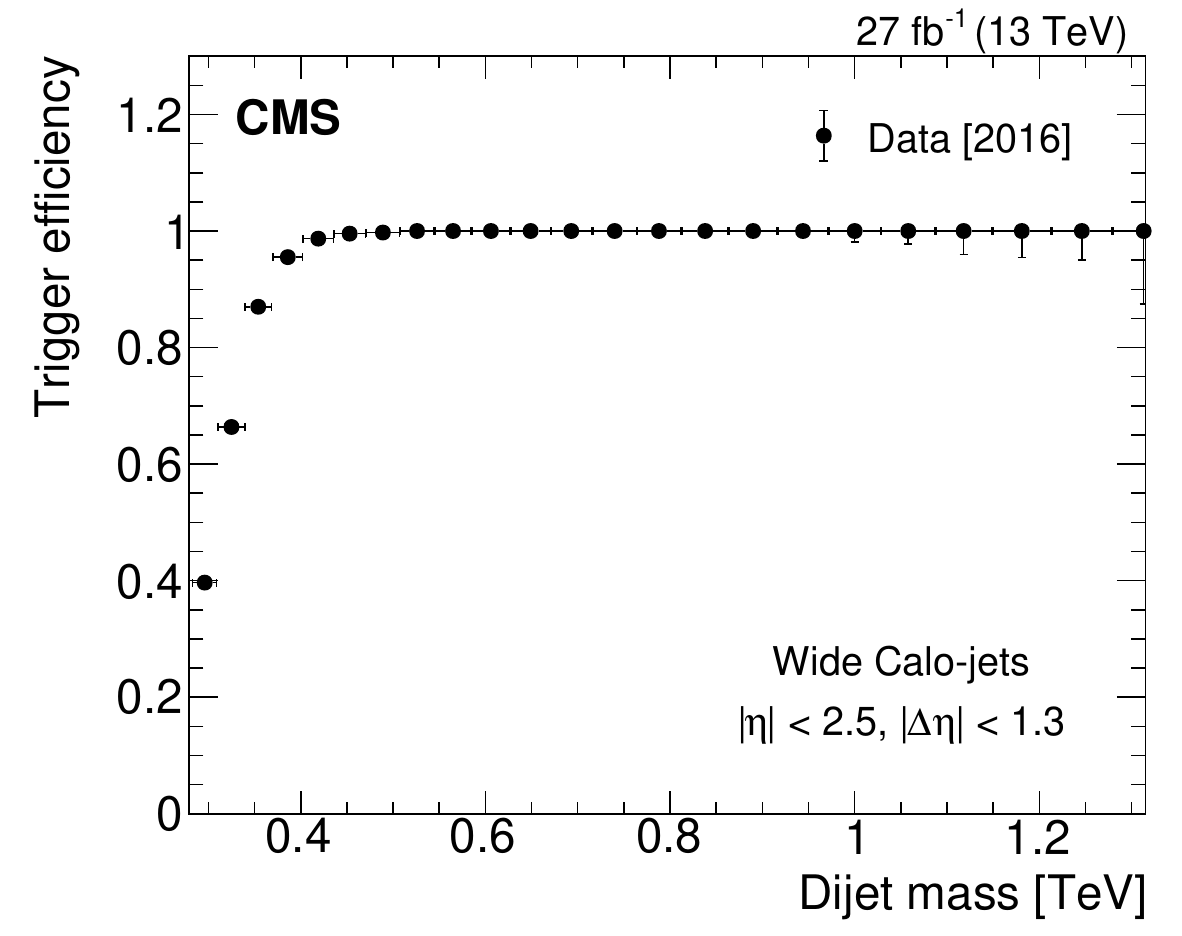}
\includegraphics[width=0.32\textwidth]{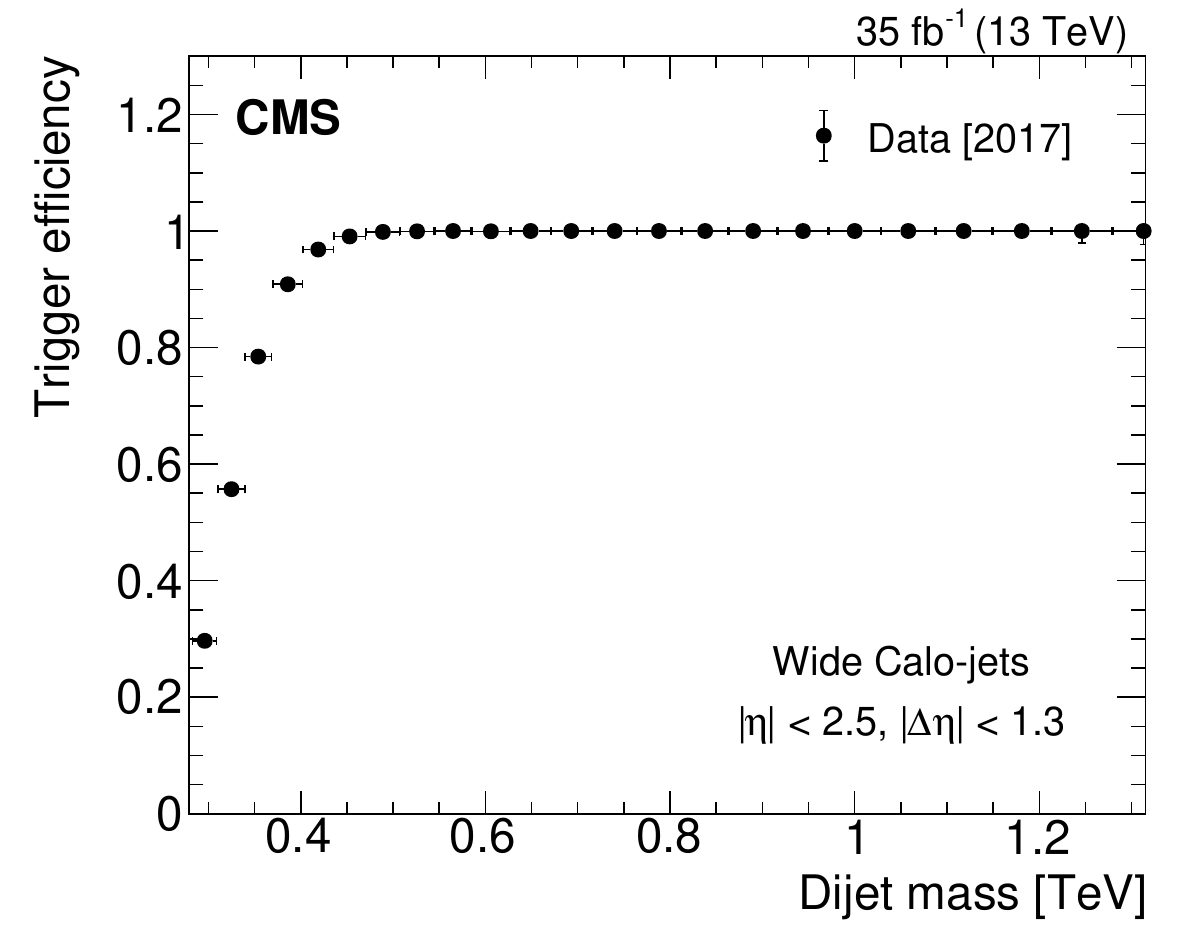}
\includegraphics[width=0.32\textwidth]{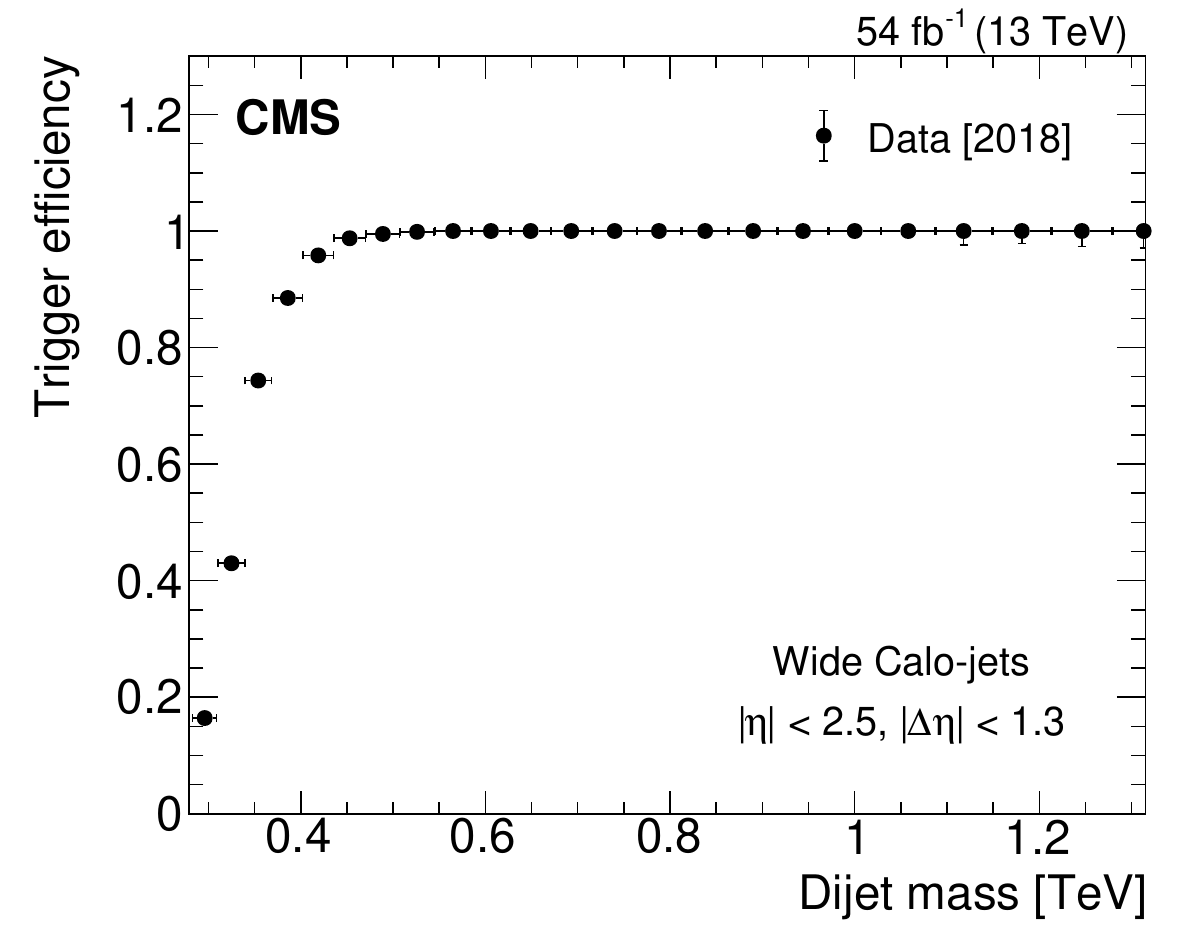}
\caption{The measured HLT trigger efficiency as a function of the offline dijet mass for wide Calo-jets, defined in Section~\ref{sec:eventselection}, for 2016 (left), 2017 (middle), and 2018 (right) data.}
\label{fig:TriggerEff}
\end{figure*}

Two auxiliary scouting trigger paths are used to measure the performance of the main \HT trigger. A prescaled L1-only trigger is used to measure the \HT trigger efficiency as a function of the dijet mass. This prescaled trigger selects a subset of the events passing the L1 \HT requirement without reference to the HLT. An even looser trigger is used to measure the efficiency of the L1 \HT seeds. This trigger is seeded by the zero-bias L1 algorithm and requires the presence of at least one Calo-jet at the HLT level. Figure~\ref{fig:TriggerEff} shows the efficiency of the trigger for the 2016, 2017, and 2018 data, demonstrating that the search is fully efficient for events with $\mjj\gtrsim\CALOminMjjCut$.

\section{Event selection}
\label{sec:eventselection}

The jet momenta and energies are corrected using calibration constants obtained from simulation, results from dedicated calibration campaigns with beams, and $\Pp\Pp$ collisions at $\sqrt{s}=13\TeV$ using the methods described in Ref.~\cite{Khachatryan:2016kdb}. Jet identification (ID) criteria are applied to remove spurious jets associated with calorimeter noise~\cite{CMS:2020ebo,CMS:2017wyc}. The jet ID requires that the jet is detected by both the ECAL and hadron calorimeters, with the fraction 5-95\% of jet energy deposited within the ECAL. Jets are required to have $\pt>30\GeV$, $\abs{\eta}<2.5$, and to satisfy the jet ID criteria. The two jets with the highest \pt are defined as the leading jets.

Spatially close jets are combined into ``wide jets'' and used to determine the dijet mass, as in the previous CMS searches~\cite{Chatrchyan2011123,CMS:2012yf,Chatrchyan:2013qhXX,Khachatryan:2015sja,Khachatryan:2015dcf,Khachatryan:2016ecr,Sirunyan:2016iap}. The two leading jets are used as seeds, and the four-vectors of all other jets, if within $\Delta R=\sqrt{\smash[b]{(\Delta\eta)^2 + (\Delta\phi)^2}}<1.1$ and satisfying $\pt>30\GeV$, are added to the nearest leading jet to obtain two wide jets, which then form the dijet system. The wide-jet algorithm thereby collects hard gluon radiation, found near the leading two final-state partons, in order to improve the dijet mass resolution. The dijet mass is the invariant mass of the two wide jets, the square root of the difference of the squared values of the total energy, $E$, and total momentum vector, $\vec{p}$, of the dijet system: $\mjj=[(E_{1}+E_{2})^2-(\vec{p_{1}}+\vec{p_{2}})^2]^{1/2}$.

{\tolerance=1800
The background from quantum chromodynamics (QCD) $t$-channel dijet events has the same angular distribution as the Rutherford scattering, approximately proportional to $1/[1-\tanh(\detajj/2)]^2$, which peaks at large values of \detajj and is suppressed by requiring the pseudorapidity of the two wide jets to satisfy $\detajj<1.3$. The above requirements maximize the search sensitivity for isotropic decays of dijet resonances in the presence of QCD dijet background. In addition, the above requirement makes the trigger efficiency, presented in Fig.~\ref{fig:TriggerEff}, reaching 100\% for relatively low values of dijet mass. This is because the jet \pt threshold of the trigger at a fixed dijet mass is more easily satisfied at low \detajj, as seen by the approximate relation $\mjj\approx 2\pt\cosh(\detajj/2)$. The Calo-jet energy corrections for the 2016 data were complete, derived from in-situ measurements of the jet response, in contrast to the Calo-jet corrections for 2017 and 2018 data, which were derived from simulation and not verified with data. Therefore, after all the jet corrections and event selections, we apply scale factors to the differential cross sections as a function of dijet mass observed in 2017 and 2018 to ensure consistency with that in the 2016 data.
\par}

\section{Signal and background modeling}
\label{sec:signalandfit}

\begin{figure}[htbp]
\centering
\includegraphics[width=0.65\textwidth]{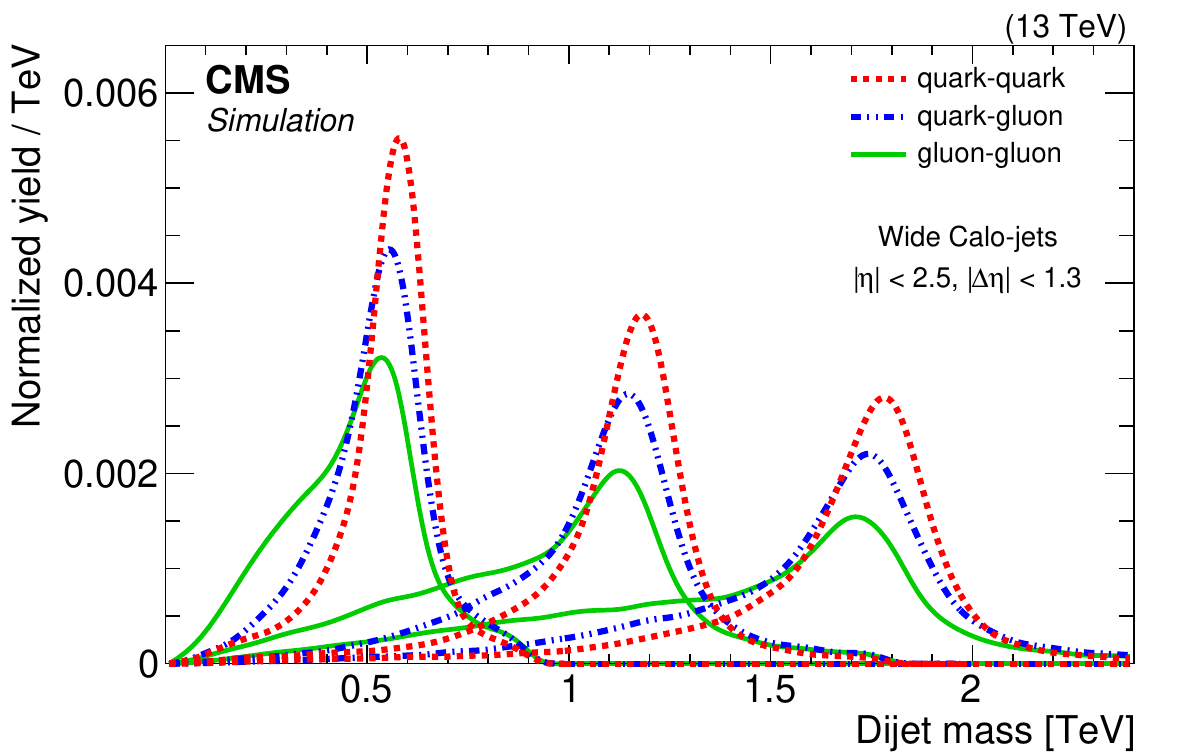}
\caption{Simulated signal shapes of narrow resonances from parton pairs quark-quark (dotted red curves), quark-gluon (dashed-dotted blue curves), and gluon-gluon (solid green curves) with masses of 0.6, 1.2, and 1.8\TeV. The reconstructed dijet mass spectra are for wide Calo-jets.}
\label{fig:signalShapes}
\end{figure}

We use independent narrow-resonance shapes for each generic type of final state: $\PQq\PQq$, $\PQq\Pg$, and $\Pg\Pg$. Resonance shapes for these three final states are shown in Fig.~\ref{fig:signalShapes}, for narrow resonances, generated with the \PYTHIA8.205~\cite{Sjostrand:2014zea} Monte Carlo (MC) program with the CUETP8M1 tune~\cite{Skands:2014pea,Khachatryan:2015pea} and including a \GEANTfour-based~\cite{refGEANT} simulation of the CMS detector for scouting reconstruction. The $\PQq\PQq$ resonances are modeled by $\PQq\PAQq\to\PXXG\to\PQq\PAQq$, the $\PQq\Pg$ resonances are modeled by $\PQq\Pg\to\Qstar\to\PQq\Pg$, and the $\Pg\Pg$ resonances are modeled by $\Pg\Pg\to\PXXG\to\Pg\Pg$, where \PXXG is the Randall--Sundrum graviton~\cite{ref_rsg} and \Qstar is an excited quark~\cite{ref_qstar,Baur:1989kv}. The predicted mass distributions have Gaussian cores from jet energy resolution (JER) and low-mass tails from QCD radiation. The observed width and low-mass tail of the resonance depend on its parton content ($\PQq\PQq$, $\PQq\Pg$, or $\Pg\Pg$). Resonances decaying to gluons, which emit more QCD radiation than quarks, are broader and have a more pronounced tail. The dijet mass resolution within the Gaussian core of $\Pg\Pg$ ($\PQq\PQq$) resonances in Fig.~\ref{fig:signalShapes} varies from 14.0 (10.5)\% at a resonance mass of 0.6 \TeV to 8.2 (6.7)\% at that mass of 1.8 \TeV.

\clearpage

All final-state parton signal resonance shapes in 2016, 2017, and 2018 MC simulation agree with each other to within 1--2\%, hence, the 2016 shapes are used to set limits and evaluate signal significances for all years.

The background estimation is obtained directly from data, fitting the monotonically and smoothly falling dijet mass spectrum with empirical parametric functions that have been extensively used in previous searches~\cite{ATLAS2010,Khachatryan:2010jd,Aad:2011aj,Harris:2011bh,Chatrchyan2011123,Aad201237,ATLAS:2012pu,CMS:2012yf,Chatrchyan:2013qhXX,Aad:2014aqa,Khachatryan:2015sja,ATLAS:2015nsi,Khachatryan:2015dcf,Khachatryan:2016ecr,Sirunyan:2016iap,Aaboud:2017yvp,Sirunyan:2018xlo,Aad:2019hjw,Sirunyan:2019vgj,CMS:2022usq,ATLAS2025Dijet}. Several functions can be considered for this background fit, organized in ``families'' with increasing numbers of parameters. The fit minimizes the negative log-likelihood function ($\Delta$NLL) of the data and the fitting function. The criteria to choose the optimal fitting function are two-fold. First, the fit must provide a good description of the data with high fit probabilities for all data sets. Second, it must produce stable, continuous $\Delta$NLL curves as a function of the signal strength for each tested resonance mass, as described in the next Section, keeping correlations between the signal shapes and the background function parameters at a minimum. The function that satisfied the above criteria and provided the best description of all different data sets is the modified exponential function with four parameters: $p_{0} \exp(p_{1} x^{p_{2}} + p_{1} (1 - x)^{p_{3}})$, where $x=\mjj/\sqrt{s}$ is the independent variable, and the $p_{i}$'s represent the fit parameters. This fitting approach yielded stable results with increased sensitivity compared to the previous CMS search~\cite{Sirunyan:2018xlo}. The improved accuracy of our background modeling can be attributed to both the use of a more effective functional form for describing the data and the need for fewer parameters to satisfy Fisher's F-test~\cite{Fisher1922}.

Signal injection tests were performed to investigate the potential bias introduced through the choice of background parameterization. As an alternative choice of parameterization we used $p_{0}(1 - x)^{p_1}x^{-(p_2 + p_3 \ln(x))}$. Pseudo-data were generated, assuming the presence of $\PQq\PQq$, $\PQq\Pg$, and $\Pg\Pg$ dijet resonance signals and the alternative parameterization of the background, and then were fit with the nominal parameterization. For all tested resonance masses, final states, background parameterizations, and injected signals, the resulting bias in the extracted signal strength is much smaller than the combined statistical and systematic uncertainty, and therefore, has a negligible impact on the upper limits.

The high statistical precision of the dijet mass spectra made it difficult to get proper fits to the complete Run-2 data sample, or even the individual data-taking years, and to achieve smooth, continuous $\Delta$NLL curves. Parametric background functions could properly describe the spectra only when extra free parameters were introduced, yet these additional free parameters created correlations with the signal shapes and generated discontinuous $\Delta$NLL curves. To alleviate this problem, hence ensuring high goodness-of-fit and consistent behavior across the fitted parameter space, we simultaneously analyzed multiple data samples, each sample with fewer events but the same total events for all samples combined. These samples were the pre-existing ``eras" of each year (2016, 2017, 2018) corresponding to different data-taking periods with slightly different running conditions: there were six periods in 2016, four in 2017, and four in 2018. The dijet spectra in these fourteen periods were then fit simultaneously to both the signal and background hypotheses, each yielding individually good fits and producing consistent and well-behaved $\Delta$NLL curves.

\begin{figure*}[!ht]
\centering
\includegraphics[width=0.435\textwidth]{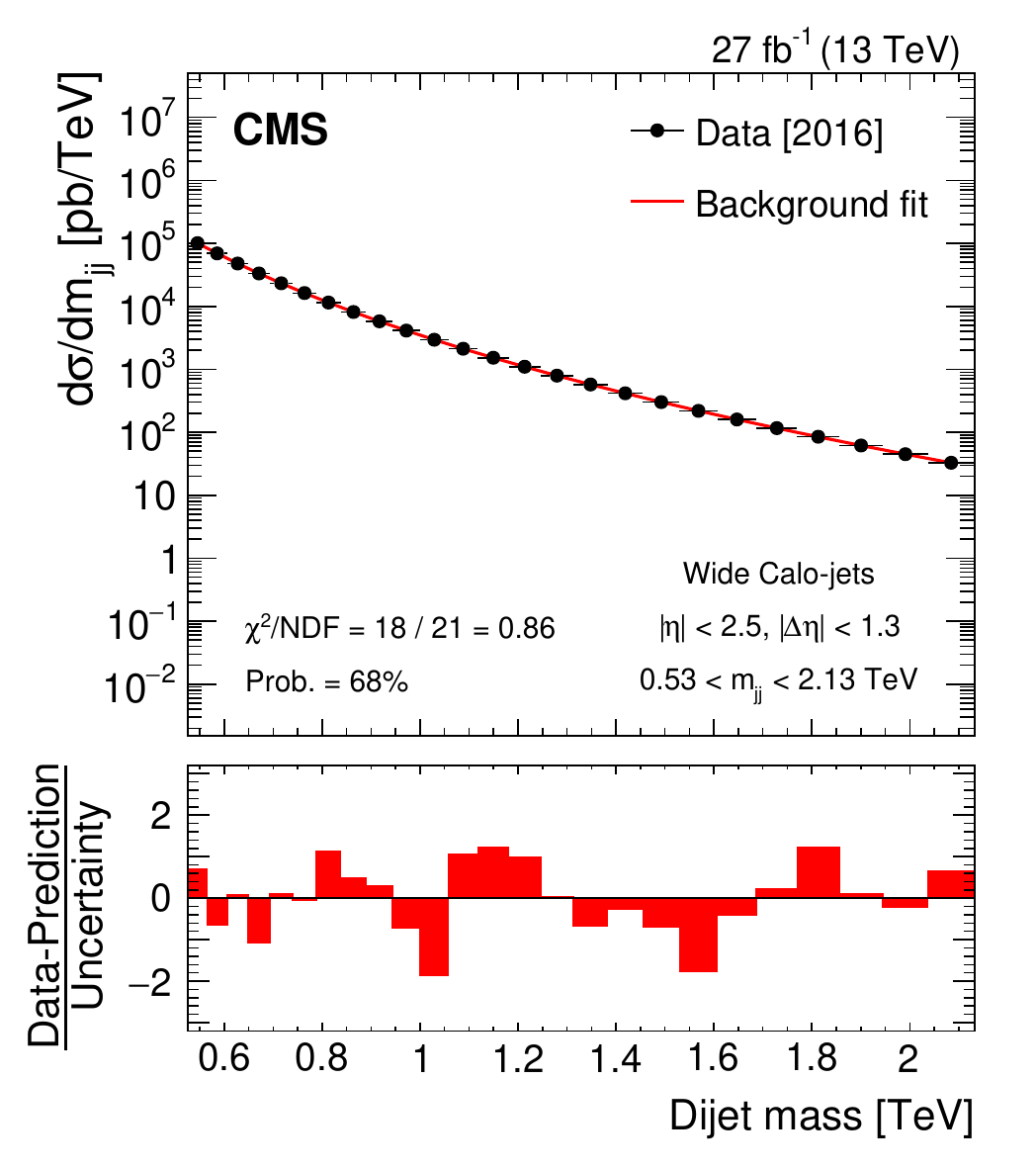}
\includegraphics[width=0.435\textwidth]{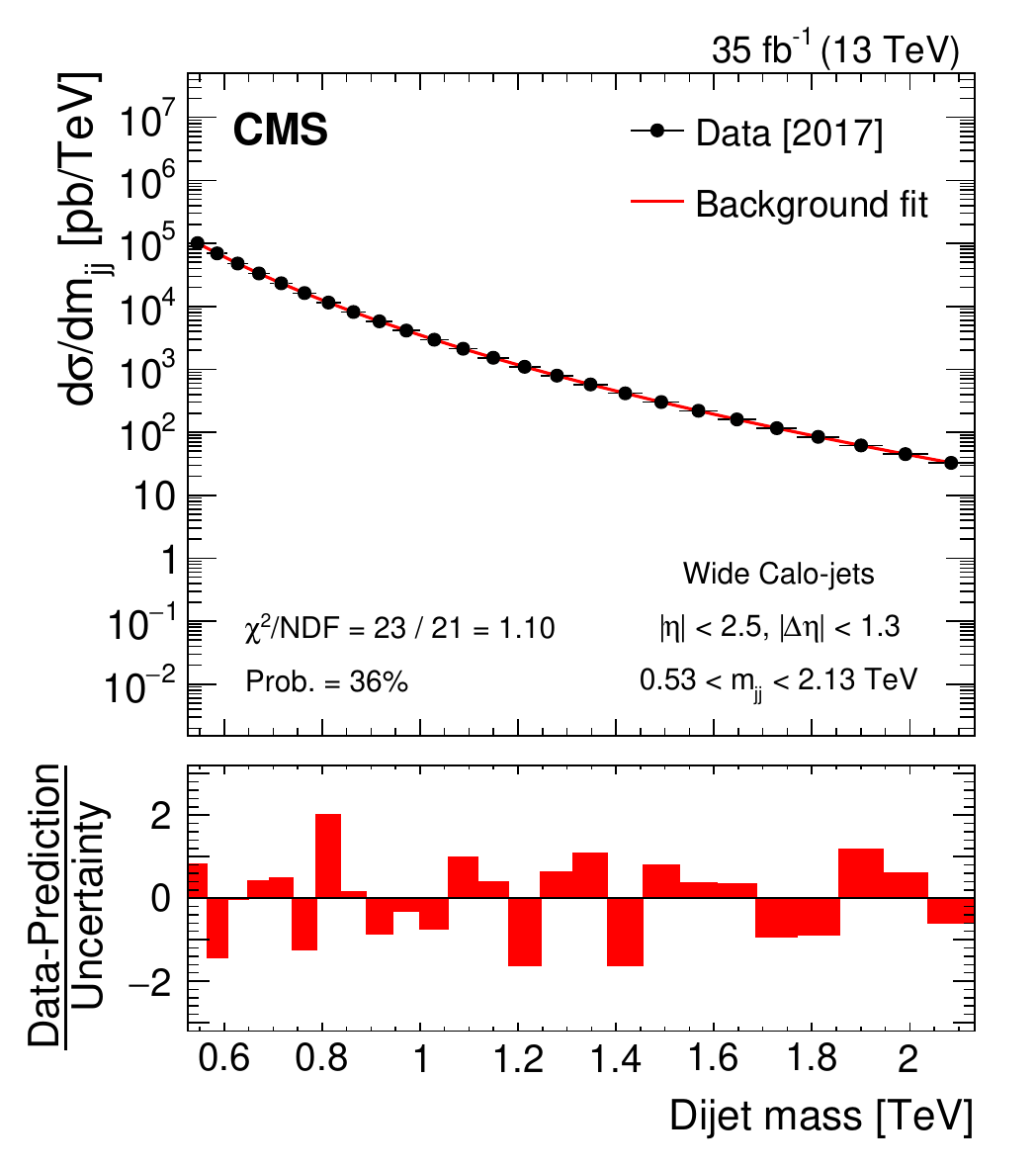} \\
\includegraphics[width=0.435\textwidth]{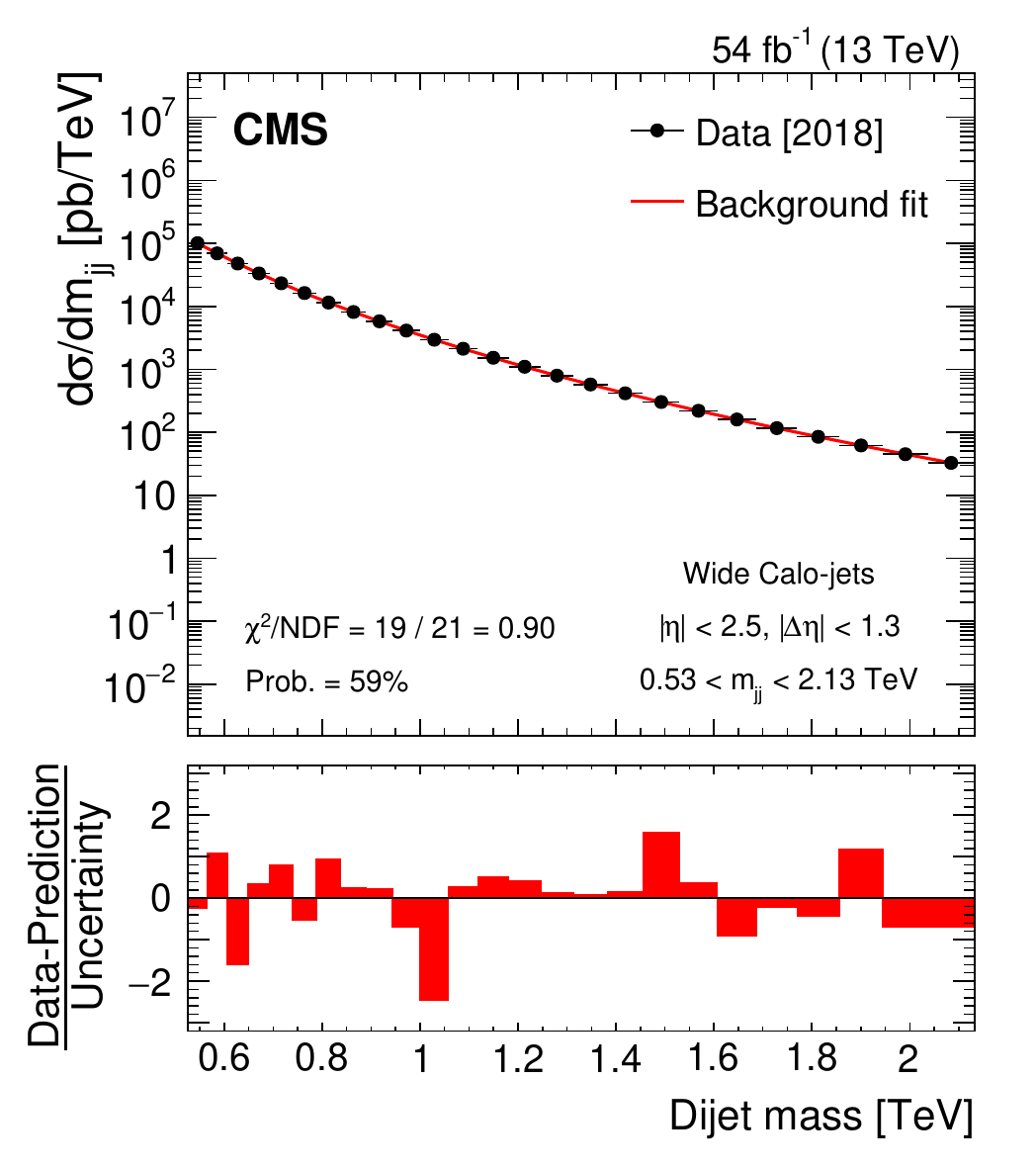}
\includegraphics[width=0.435\textwidth]{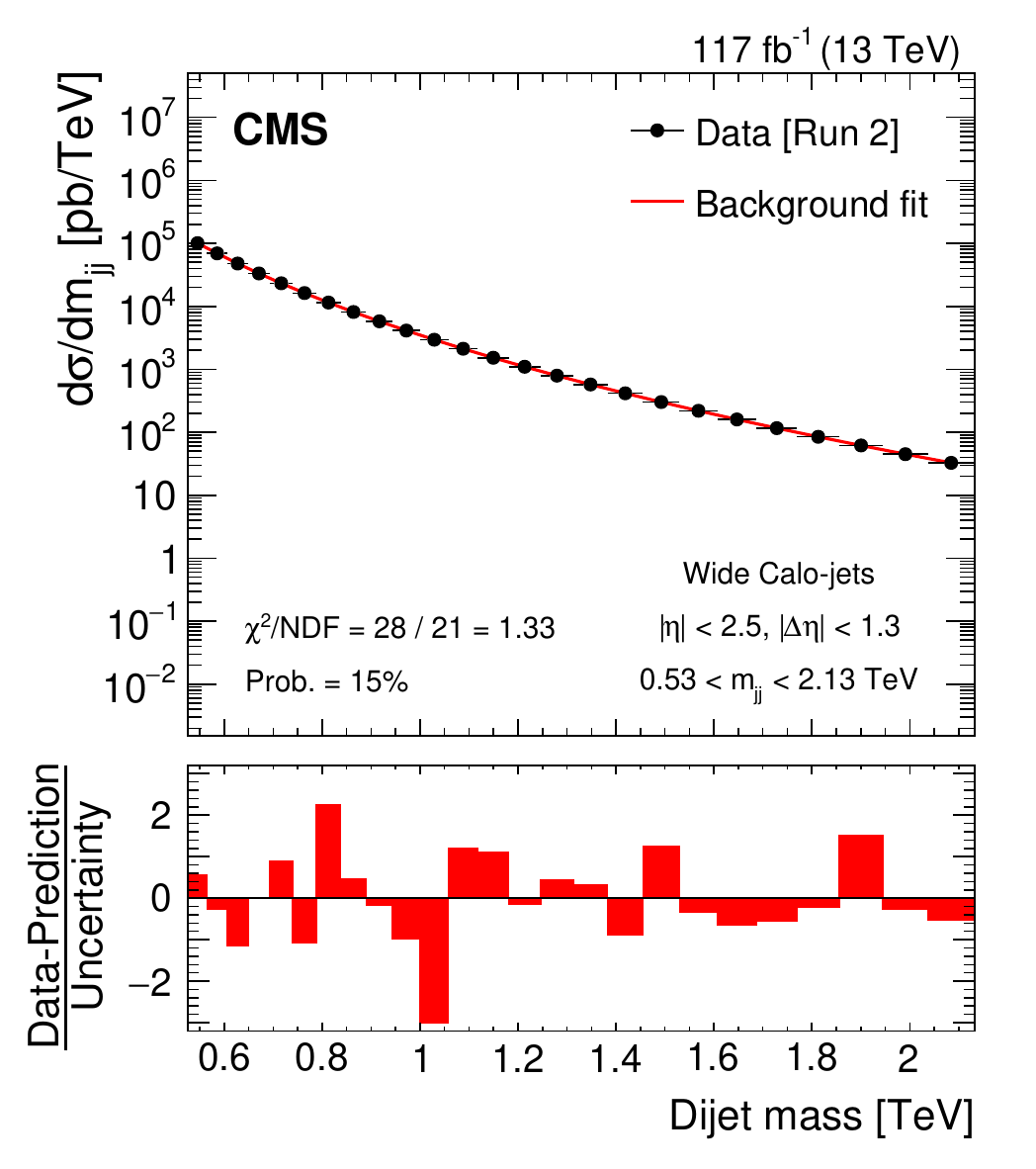}
\caption{Dijet mass spectra for wide Calo-jets (points) compared to a parameterization of the background (solid curve) for the 2016 (upper left), 2017 (upper right), 2018 (lower left), and the combined (lower right) data sets. The horizontal lines on the data points in the upper panel show variable bin sizes, while the red bars in the lower panels show the bin-by-bin difference between the data and parameterization, normalized to the total uncertainty.}
\label{fig:FullDatasetFitResultsCalibrated}
\end{figure*}

Figure~\ref{fig:FullDatasetFitResultsCalibrated} shows the djiet mass spectra from each year, 2016, 2017, 2018, and the sum of all three years (Run-2), along with their background-only fits. The binning of the spectra is based on variable-width bins that increase with \mjj and are chosen to be roughly comparable to the dijet mass resolution over the full mass range. The background curves shown are the integrated luminosity weighted sum of the relevant background-only fits from the fourteen data-taking periods within the three years. Also shown are the pulls, defined as the difference between the data and the fitted parameterization, divided by the statistical uncertainty of the data. The dijet mass spectra are well modeled by the background fits, as evidenced by the fit probability and the pulls, and there is no significant evidence for resonant particle production.

\section{Statistical procedure}
\label{sec:fitmethod}

To set limits, we use a multibin counting experiment likelihood, which is a product of Poisson distributions corresponding to different bins. The sources of systematic uncertainty considered are the jet energy scale (JES) and JER, the integrated luminosity, and the values of the parameters within the functional form modeling the background shape in the dijet mass distribution. The background systematic uncertainties are related to the parameters of the functional forms utilized, that were left unconstrained in the fit, and are by far the dominant ones in these searches. The uncertainty in the JES is within \jecUncert for all values of the dijet mass and is determined from $\sqrt{s}=13\TeV$ data using the methods described in Ref.~\cite{Khachatryan:2016kdb}. This uncertainty is propagated to the limits by shifting the dijet mass shape for the signal by $\pm$\jecUncert. The uncertainty in the JER translates into an uncertainty of 10\% in the resolution of the dijet mass~\cite{Khachatryan:2016kdb}, and it is propagated to the limits by observing the effect of increasing and decreasing by 10\% the reconstructed width of the dijet mass shape for the signal. The uncertainties in the integrated luminosities are 1.2\% in 2016~\cite{Sirunyan:2759951}, 2.3\% in 2017~\cite{CMS-PAS-LUM-17-004}, and 2.5\% in 2018~\cite{CMS-PAS-LUM-18-002}, and are propagated to the signal normalization.

The test statistic is defined following the LHC \CLs criterion~\cite{Junk:1999kv,Read:2002hq} and is implemented using the CMS statistical analysis tool \textsc{Combine}~\cite{CMS:2024onh}, which is based on the \textsc{RooFit}~\cite{Verkerke:2003ir} and \textsc{RooStats}~\cite{Moneta:2010pm} frameworks. In the frequentist paradigm, the systematic uncertainties related to the nuisance parameters are taken into account through profiling (maximization of the likelihood). The sources of systematic uncertainty are implemented as nuisance parameters in the likelihood model. For the background parameterization, parameters are considered as freely floating parameters for each data-taking period independently and uncorrelated across years. Signal modeling uncertainties on JES and JER are implemented with Gaussian constraints, and on the integrated luminosity with log-normal constraints, correlated across years. We then use this test statistic to derive the observed 95\% \CL upper limit on the signal cross section in the asymptotic approximation~\cite{Cowan:2010js} and to estimate the local significance.

\section{Limits and significance}
\label{sec:results}

\begin{figure}[!ht]
\centering
\includegraphics[width=0.49\textwidth]{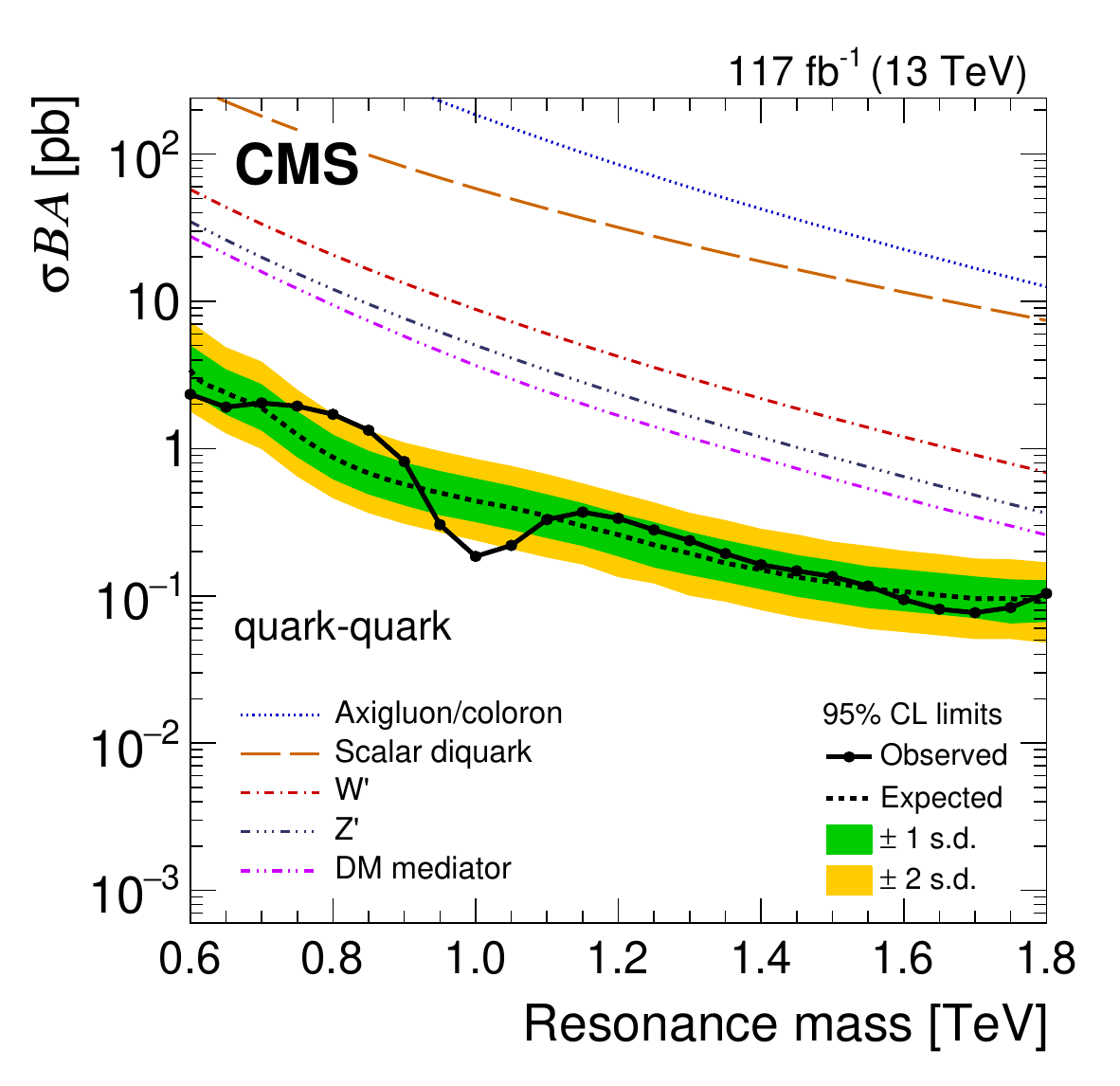}%
\includegraphics[width=0.49\textwidth]{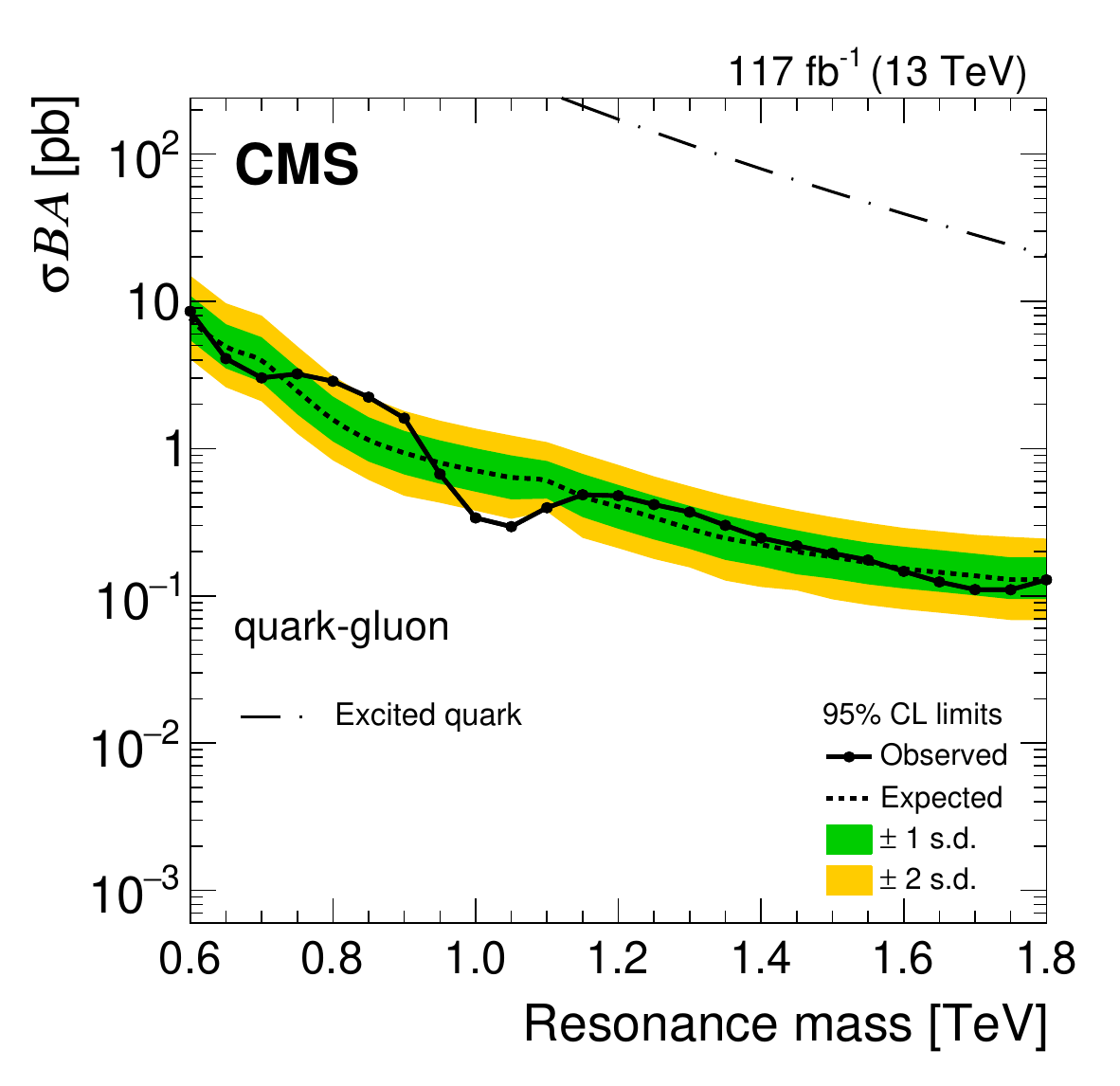}\\
\includegraphics[width=0.49\textwidth]{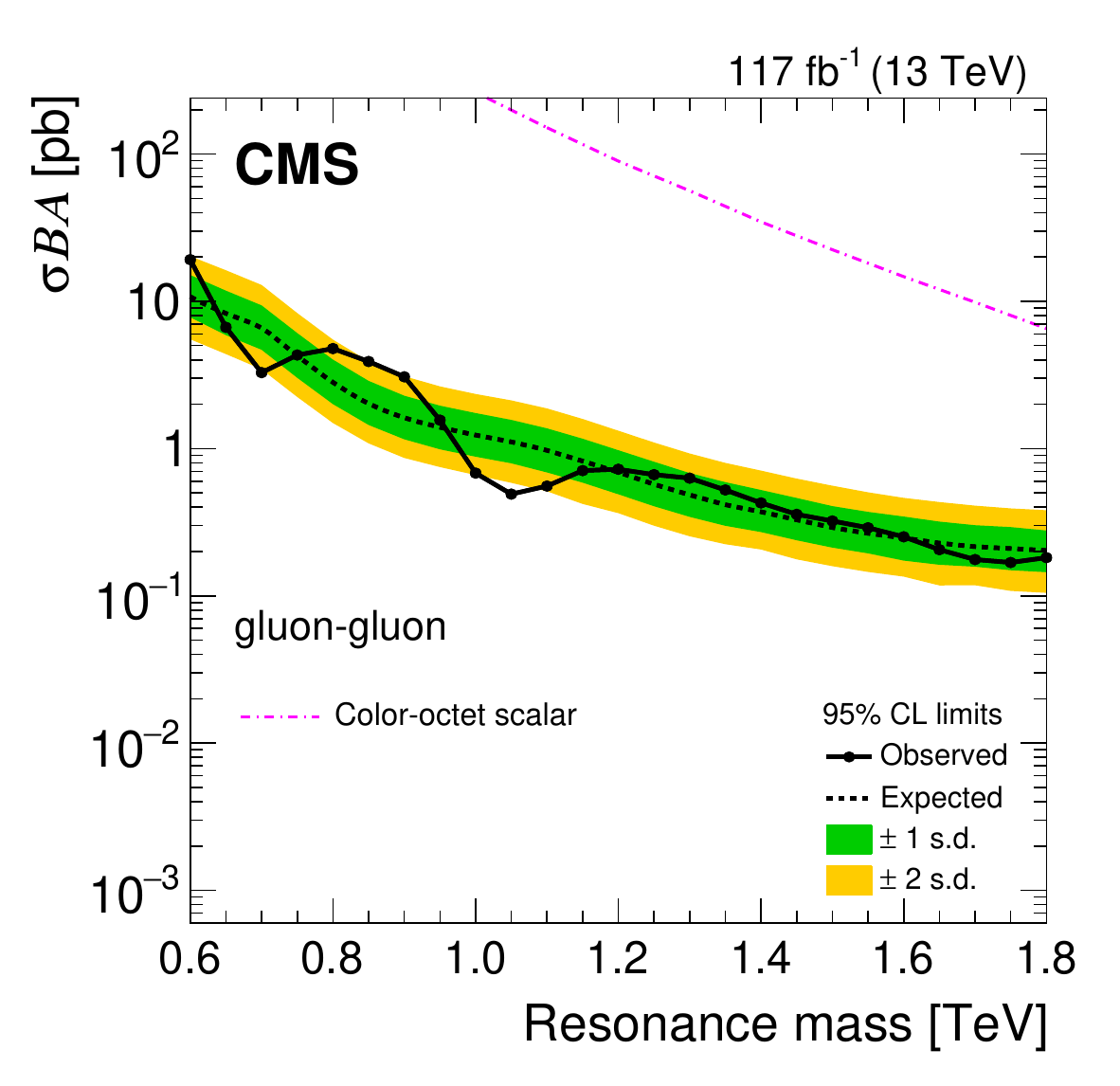}
\caption{The observed 95\% \CL upper limits on the product of the cross section ($\sigma$), branching fraction ($B$), and acceptance ($A$) for dijet resonances decaying to quark-quark (upper left), quark-gluon (upper right), and gluon-gluon (lower). The corresponding expected limits (dashed) and their variations at the 1 and 2 standard deviation levels (shaded bands) are also shown. Limits are compared to the predicted cross sections for new gauge bosons $\PWpr$ and $\PZpr$ with SM-like couplings~\cite{ref_gauge}, axigluons~\cite{ref_axi}, excited quarks~\cite{ref_qstar,Baur:1989kv}, scalar diquarks~\cite{ref_diquark}, colorons~\cite{ref_coloron}, color-octet scalars~\cite{Han:2010rf}, and DM mediators for the couplings $\gq=0.25$ and $\gDM=1$, and dark matter mass $\mDM=1\GeV$~\cite{BOVEIA2020100365,Abdallah:2015ter}.}
\label{fig:limits}
\end{figure}

We use the dijet mass spectrum from wide jets, the background parameterization, and the dijet resonance shapes to set limits on the product of the production cross section ($\sigma$), branching fraction ($B$), and acceptance ($A$), of new particles decaying to the parton pairs $\PQq\PQq$ (or $\PQq\PAQq$), $\PQq\Pg$, and $\Pg\Pg$. A separate limit is determined for each final state ($\PQq\PQq$, $\PQq\Pg$, and $\Pg\Pg$) because of the dependence of the dijet resonance shape on the types of the two final-state partons. Figure~\ref{fig:limits} shows the 95\% \CL asymptotic LHC \CLs limits on ${\sigma}{BA}$ for the three considered final states. Therefore, the acceptance of the minimum dijet mass requirement does not appear in the acceptance $A$.

All upper limits presented can be compared to the parton-level predictions of ${\sigma}{BA}$ without detector simulation, to determine mass limits on new particles. The model predictions shown in Fig.~\ref{fig:limits} are lowest-order parton-level calculations, as previously discussed~\cite{Sirunyan:2018xlo}. The acceptance of the minimum dijet mass requirement for each considered signal has been evaluated separately for each final state and has been taken into account by correcting the limits. The acceptance is evaluated at the parton level for the resonance decay to two partons. In the case of isotropic decays, the acceptance is $A\approx 0.6$ and is independent of the resonance mass. For a given model, new particles are excluded at 95\% \CL in mass regions where the theoretical prediction lies at or above the observed upper limit for the appropriate final state of Fig.~\ref{fig:limits}. Between a resonance mass of 0.6 and 1.8\TeV, we exclude all the benchmark models for the coupling values considered, as shown in Fig.~\ref{fig:limits}.

\begin{figure}[!htbp]
\includegraphics[width=0.32\textwidth]{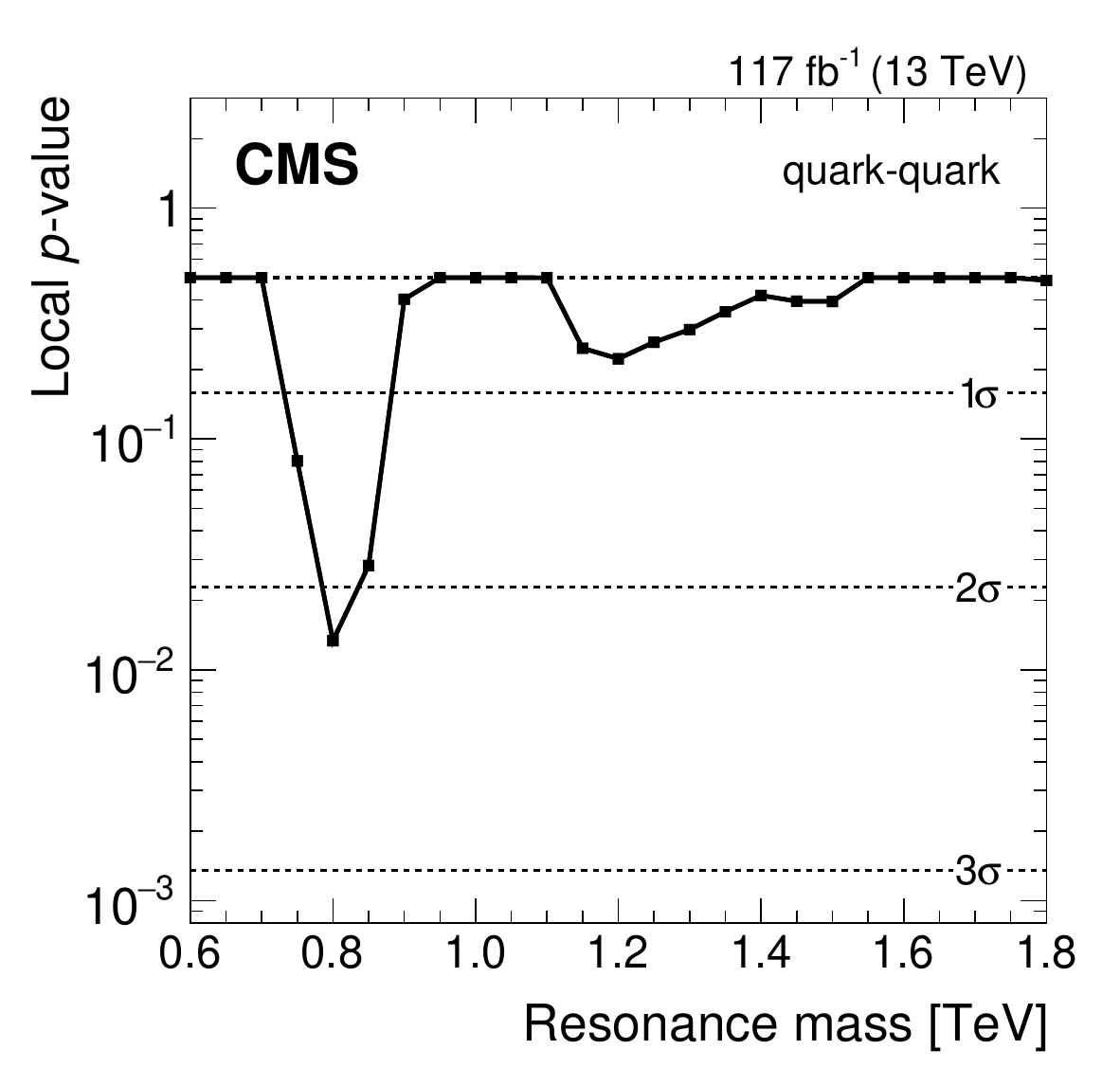}%
\includegraphics[width=0.32\textwidth]{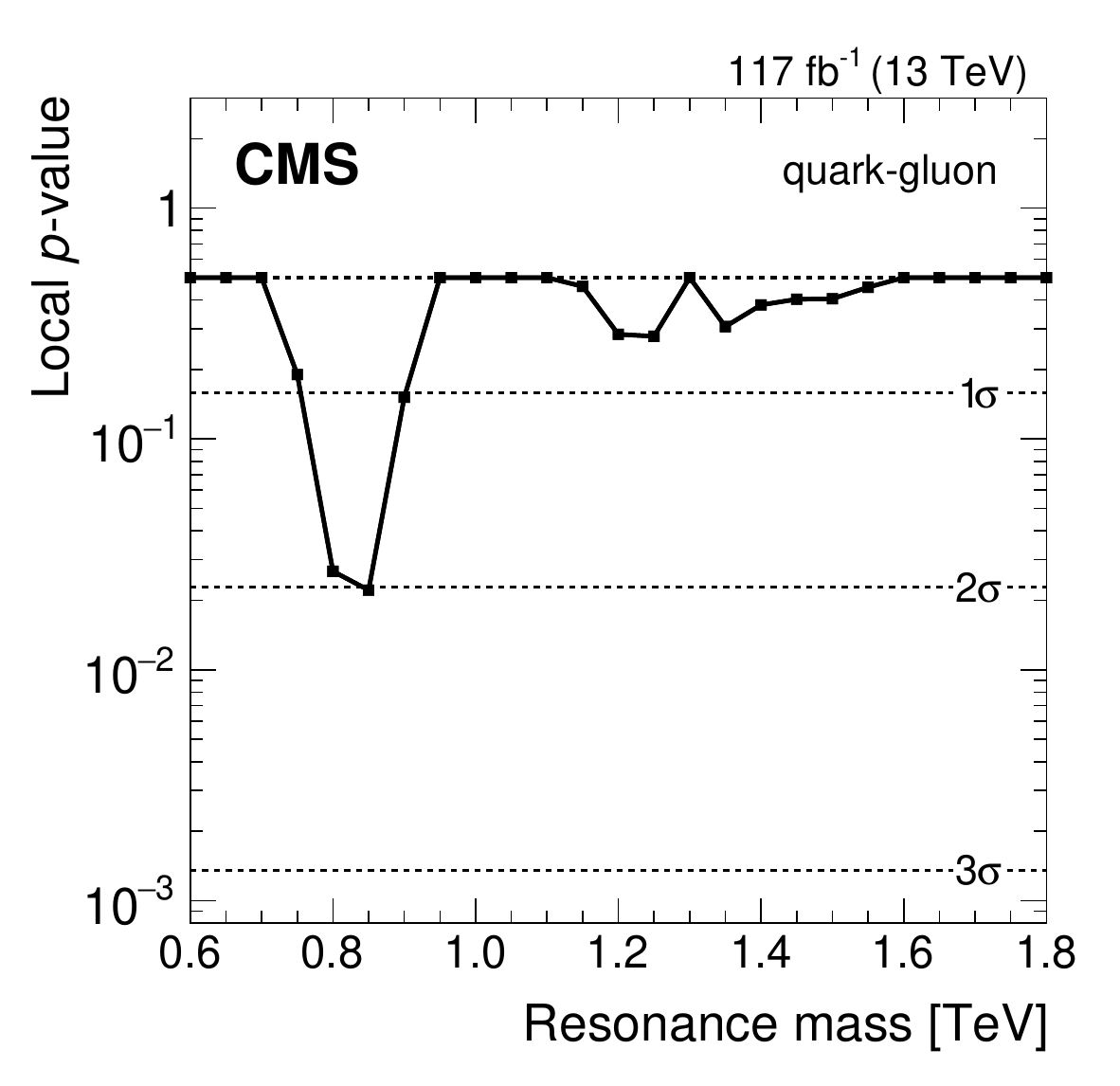}%
\includegraphics[width=0.32\textwidth]{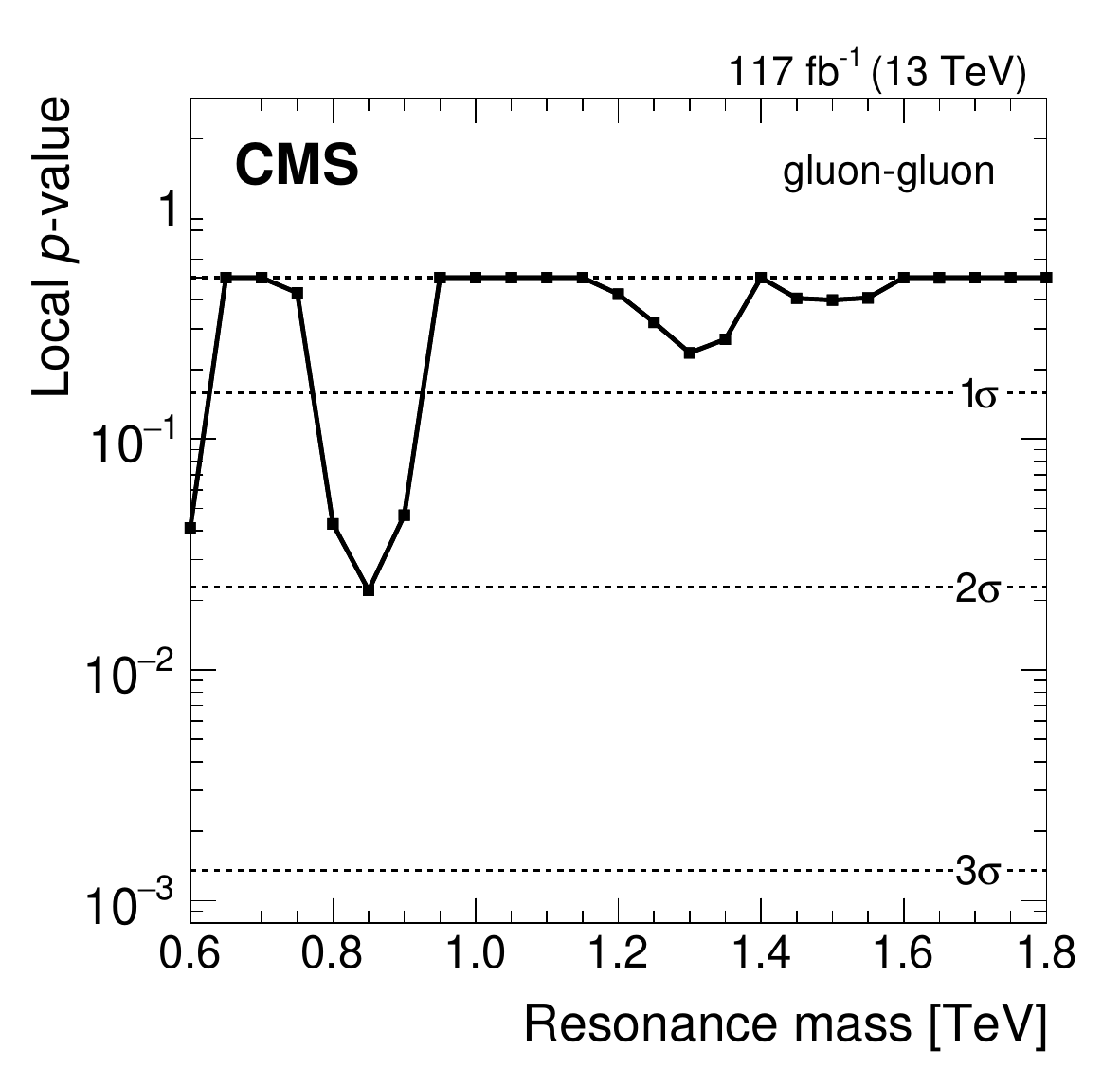}
\caption{Local \pvalue, and the corresponding significance in standard deviations, for quark-quark (left), quark-gluon (middle), and gluon-gluon (right) resonances.}
\label{fig:signif}
\end{figure}

Figure~\ref{fig:signif} shows the \pvalue, a measure of the local significance, as a function of mass for $\PQq\PQq$, $\PQq\Pg$, and $\Pg\Pg$ resonances. The highest local significances (lowest \pvalues), of roughly 2 standard deviations, are observed at resonance masses in the 0.80-0.85\TeV region in all three cases.

\section{Limits on dark-matter mediators}
\label{sec:dm_interpr}

\begin{figure}[!htbp]
\centering
\includegraphics[width=0.78\textwidth]{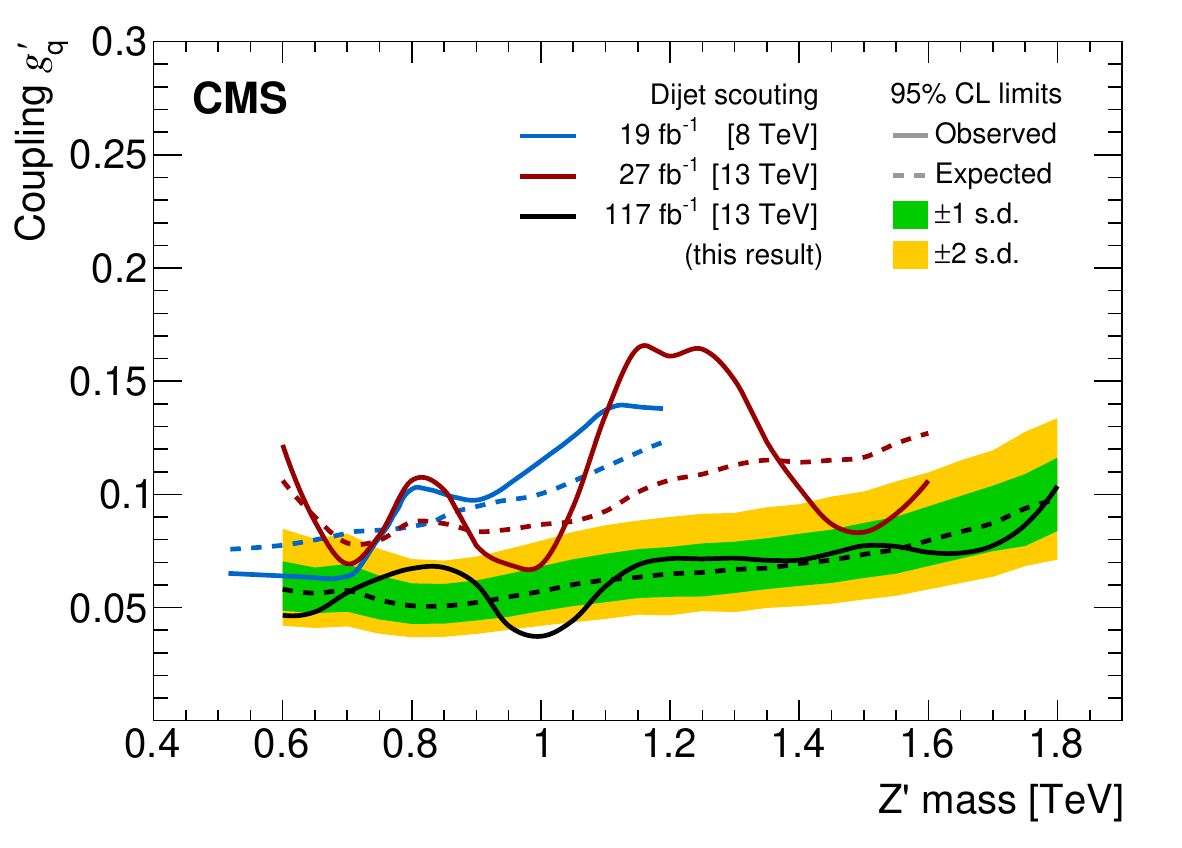}
\caption{The 95\% \CL upper limit on the universal quark coupling $\gqPrime$ as a function of resonance mass for a leptophobic $\PZpr$ resonance that only couples to quarks. The observed limits (solid), expected limits (dashed), and their variation at the 1 and 2 standard deviation levels (shaded bands) are shown. Current limits (black) are compared with previously published ones from CMS at 8\TeV~\cite{Khachatryan:2016ecr} (blue) and 13\TeV~\cite{Sirunyan:2018xlo} (red).}
\label{fig:dm}
\end{figure}

Figure~\ref{fig:dm} presents the excluded values of the universal quark coupling as a function of the $\PZpr$ boson mass. It shows the significant improvement of the current limits in the coupling versus mass plane with respect to the previously published results from CMS at 8\TeV~\cite{Khachatryan:2016ecr} and 13\TeV~\cite{Sirunyan:2018xlo}. We note that the improvement is beyond what is expected for the statistical component of the limit, which usually scales inversely proportional to the square root of the integrated luminosity. This is a consequence of the new statistical approach, which requires fewer parameters in the empirical functional forms of the background modeling. For the first time, CMS demonstrates sensitivity to coupling values as small as 0.04. Our sensitivity is similar to that from the most recent search by the ATLAS Collaboration~\cite{ATLAS2025Dijet} based on an integrated luminosity of 132\fbinv. The previous ATLAS search~\cite{ATLAS:2018qto}, although it used much less integrated luminosity, reported significantly higher sensitivity than the recent CMS and ATLAS searches for $\PZpr$ masses less than approximately 1\TeV, and has been explicitly superseded by the most recent ATLAS publication~\cite{ATLAS2025Dijet}. In contrast to the prior ATLAS publication, their most recent paper estimates the important uncertainty in the background in the same way as CMS does, by treating the parameters of the background functional form as nuisances and allowing them to float freely when maximizing the likelihood.

\clearpage

\section{Summary}
\label{sec:conclusions}

Calorimeter jets from data scouting have been used to search for dijet resonances with masses between 0.6 and 1.8\TeV. The dijet mass spectra are observed to be smoothly falling distributions, and no significant evidence for resonant particle production is found. Signal significances and upper limits are presented as functions of the resonance mass, on the product of the cross section, branching fraction, and acceptance, for the individual cases of narrow quark-quark, quark-gluon, and gluon-gluon resonances that are applicable to any model of narrow dijet resonance production. The largest local significance is observed to be 2.2 standard deviations at a quark-quark resonance mass of 0.8\TeV. The limits exclude models of color-octet scalars, excited quarks, axigluons, colorons, scalar diquarks, $\PWpr$ and $\PZpr$ bosons, and dark matter (DM) mediators, for the benchmark choices of the couplings, over the complete mass range considered. The limit on the coupling of DM mediators to quarks, in a simplified model of interactions between quarks and DM, is presented as a function of the mediator mass. The sensitivity of this search goes beyond what is expected from statistical scaling with the integrated luminosity alone, as a consequence of the use of fewer parameters in the background function within a more robust statistical procedure.

\begin{acknowledgments}
We congratulate our colleagues in the CERN accelerator departments for the excellent performance of the LHC and thank the technical and administrative staffs at CERN and at other CMS institutes for their contributions to the success of the CMS effort. In addition, we gratefully acknowledge the computing centers and personnel of the Worldwide LHC Computing Grid and other centers for delivering so effectively the computing infrastructure essential to our analyses. Finally, we acknowledge the enduring support for the construction and operation of the LHC, the CMS detector, and the supporting computing infrastructure provided by the following funding agencies: SC (Armenia), BMBWF and FWF (Austria); FNRS and FWO (Belgium); CNPq, CAPES, FAPERJ, FAPERGS, and FAPESP (Brazil); MES and BNSF (Bulgaria); CERN; CAS, MoST, and NSFC (China); MINCIENCIAS (Colombia); MSES and CSF (Croatia); RIF (Cyprus); SENESCYT (Ecuador); ERC PRG, TARISTU24-TK10 and MoER TK202 (Estonia); Academy of Finland, MEC, and HIP (Finland); CEA and CNRS/IN2P3 (France); SRNSF (Georgia); BMFTR, DFG, and HGF (Germany); GSRI (Greece); NKFIH (Hungary); DAE and DST (India); IPM (Iran); SFI (Ireland); INFN (Italy); MSIT and NRF (Republic of Korea); MES (Latvia); LMTLT (Lithuania); MOE and UM (Malaysia); BUAP, CINVESTAV, CONACYT, LNS, SEP, and UASLP-FAI (Mexico); MOS (Montenegro); MBIE (New Zealand); PAEC (Pakistan); MES, NSC, and NAWA (Poland); FCT (Portugal);  MESTD (Serbia); MICIU/AEI and PCTI (Spain); MOSTR (Sri Lanka); Swiss Funding Agencies (Switzerland); MST (Taipei); MHESI (Thailand); TUBITAK and TENMAK (T\"{u}rkiye); NASU (Ukraine); STFC (United Kingdom); DOE and NSF (USA).

\hyphenation{Rachada-pisek} Individuals have received support from the Marie-Curie program and the European Research Council and Horizon 2020 Grant, contract Nos.\ 675440, 724704, 752730, 758316, 765710, 824093, 101115353, 101002207, 101001205, and COST Action CA16108 (European Union); the Leventis Foundation; the Alfred P.\ Sloan Foundation; the Alexander von Humboldt Foundation; the Science Committee, project no. 22rl-037 (Armenia); the Fonds pour la Formation \`a la Recherche dans l'Industrie et dans l'Agriculture (FRIA) and Fonds voor Wetenschappelijk Onderzoek contract No. 1228724N (Belgium); the Beijing Municipal Science \& Technology Commission, No. Z191100007219010, the Fundamental Research Funds for the Central Universities, the Ministry of Science and Technology of China under Grant No. 2023YFA1605804, the Natural Science Foundation of China under Grant No. 12061141002, 12535004, and USTC Research Funds of the Double First-Class Initiative No.\ YD2030002017 (China); the Ministry of Education, Youth and Sports (MEYS) of the Czech Republic; the Shota Rustaveli National Science Foundation, grant FR-22-985 (Georgia); the Deutsche Forschungsgemeinschaft (DFG), among others, under Germany's Excellence Strategy -- EXC 2121 ``Quantum Universe" -- 390833306, and under project number 400140256 - GRK2497; the Hellenic Foundation for Research and Innovation (HFRI), Project Number 2288 (Greece); the Hungarian Academy of Sciences, the New National Excellence Program - \'UNKP, the NKFIH research grants K 131991, K 133046, K 138136, K 143460, K 143477, K 146913, K 146914, K 147048, 2020-2.2.1-ED-2021-00181, TKP2021-NKTA-64, and 2021-4.1.2-NEMZ\_KI-2024-00036 (Hungary); the Council of Science and Industrial Research, India; ICSC -- National Research Center for High Performance Computing, Big Data and Quantum Computing, FAIR -- Future Artificial Intelligence Research, and CUP I53D23001070006 (Mission 4 Component 1), funded by the NextGenerationEU program (Italy); the Latvian Council of Science; the Ministry of Education and Science, project no. 2022/WK/14, and the National Science Center, contracts Opus 2021/41/B/ST2/01369, 2021/43/B/ST2/01552, 2023/49/B/ST2/03273, and the NAWA contract BPN/PPO/2021/1/00011 (Poland); the Funda\c{c}\~ao para a Ci\^encia e a Tecnologia, grant CEECIND/01334/2018 (Portugal); the National Priorities Research Program by Qatar National Research Fund;MICIU/AEI/10.13039/501100011033, ERDF/EU, "European Union NextGenerationEU/PRTR", and Programa Severo Ochoa del Principado de Asturias (Spain); the Chulalongkorn Academic into Its 2nd Century Project Advancement Project, the National Science, Research and Innovation Fund program IND\_FF\_68\_369\_2300\_097, and the Program Management Unit for Human Resources \& Institutional Development, Research and Innovation, grant B39G680009 (Thailand); the Kavli Foundation; the Nvidia Corporation; the SuperMicro Corporation; the Welch Foundation, contract C-1845; and the Weston Havens Foundation (USA).
\end{acknowledgments}
\bibliography{auto_generated}

\providecommand{\href}[2]{#2}\begingroup\raggedright\begin{thebibliography}{10}%
\makeatletter
\providecommand{\hrefCMSnoop }[0]{\@secondoftwo}%
\makeatother
\providecommand{\doi}{\texttt{doi:}\begingroup \urlstyle{tt}\Url}

\bibitem{ref_gauge}
\hrefCMSnoop {}{E.~Eichten, I.~Hinchliffe, K.~D. Lane, and C.~Quigg,
  ``Supercollider physics'',} \textit{ Rev. Mod. Phys.} \textbf{ 56} (1984)
  579,
\href{http://dx.doi.org/10.1103/RevModPhys.56.579}{\doi{10.1103/RevModPhys.56.579}}.

\bibitem{ref_axi}
\hrefCMSnoop {}{P.~H. Frampton and S.~L. Glashow, ``Chiral color: An
  alternative to the standard model'',} \textit{ Phys. Lett. B} \textbf{ 190}
  (1987) 157,
\href{http://dx.doi.org/10.1016/0370-2693(87)90859-8}{\doi{10.1016/0370-2693(87)90859-8}}.

\bibitem{ref_qstar}
\hrefCMSnoop {}{U.~Baur, I.~Hinchliffe, and D.~Zeppenfeld, ``Excited quark
  production at hadron colliders'',} \textit{ Int. J. Mod. Phys. A} \textbf{
  02} (1987) 1285,
\href{http://dx.doi.org/10.1142/S0217751X87000661}{\doi{10.1142/S0217751X87000661}}.

\bibitem{ref_diquark}
\hrefCMSnoop {}{J.~L. Hewett and T.~G. Rizzo, ``Low-energy phenomenology of
  superstring-inspired {E(6)} models'',} \textit{ Phys. Rept.} \textbf{ 183}
  (1989) 193,
\href{http://dx.doi.org/10.1016/0370-1573(89)90071-9}{\doi{10.1016/0370-1573(89)90071-9}}.

\bibitem{Baur:1989kv}
\hrefCMSnoop {}{U.~Baur, M.~Spira, and P.~M. Zerwas, ``Excited quark and lepton
  production at hadron colliders'',} \textit{ Phys. Rev. D} \textbf{ 42} (1990)
  815,
\href{http://dx.doi.org/10.1103/PhysRevD.42.815}{\doi{10.1103/PhysRevD.42.815}}.

\bibitem{ref_coloron}
\hrefCMSnoop {}{E.~H. Simmons, ``Coloron phenomenology'',} \textit{ Phys. Rev.
  D} \textbf{ 55} (1997) 1678,
  \href{http://dx.doi.org/10.1103/PhysRevD.55.1678}{\doi{10.1103/PhysRevD.55.1678}},
\href{http://www.arXiv.org/abs/hep-ph/9608269}{\texttt{arXiv:hep-ph/9608269}}.

\bibitem{ref_rsg}
\hrefCMSnoop {}{L.~Randall and R.~Sundrum, ``An alternative to
  compactification'',} \textit{ Phys. Rev. Lett.} \textbf{ 83} (1999) 4690,
  \href{http://dx.doi.org/10.1103/PhysRevLett.83.4690}{\doi{10.1103/PhysRevLett.83.4690}},
\href{http://www.arXiv.org/abs/hep-th/9906064}{\texttt{arXiv:hep-th/9906064}}.

\bibitem{Cullen:2000ef}
\hrefCMSnoop {}{S.~Cullen, M.~Perelstein, and M.~E. Peskin, ``{TeV strings and
  collider probes of large extra dimensions}'',} \textit{ Phys. Rev. D}
  \textbf{ 62} (2000) 055012,
  \href{http://dx.doi.org/10.1103/PhysRevD.62.055012}{\doi{10.1103/PhysRevD.62.055012}},
\href{http://www.arXiv.org/abs/hep-ph/0001166}{\texttt{arXiv:hep-ph/0001166}}.

\bibitem{Han:2010rf}
\hrefCMSnoop {}{T.~Han, I.~Lewis, and Z.~Liu, ``Colored resonant signals at the
  {LHC}: Largest rate and simplest topology'',} \textit{ JHEP} \textbf{ 12}
  (2010) 085,
  \href{http://dx.doi.org/10.1007/JHEP12(2010)085}{\doi{10.1007/JHEP12(2010)085}},
  \href{http://www.arXiv.org/abs/1010.4309}{\texttt{arXiv:1010.4309}}.

\bibitem{Chivukula:2013xla}
\hrefCMSnoop {}{R.~S. Chivukula, E.~H. Simmons, A.~Farzinnia, and J.~Ren,
  ``Hadron collider production of massive color-octet vector bosons at
  next-to-leading order'',} \textit{ Phys. Rev. D} \textbf{ 87} (2013) 094011,
  \href{http://dx.doi.org/10.1103/PhysRevD.87.094011}{\doi{10.1103/PhysRevD.87.094011}},
\href{http://www.arXiv.org/abs/1303.1120}{\texttt{arXiv:1303.1120}}.

\bibitem{Chala:2015ama}
M.~Chala\hrefCMSnoop {}{ { et~al.}, ``Constraining dark sectors with monojets
  and dijets'',} \textit{ JHEP} \textbf{ 07} (2015) 089,
  \href{http://dx.doi.org/10.1007/JHEP07(2015)089}{\doi{10.1007/JHEP07(2015)089}},
\href{http://www.arXiv.org/abs/1503.05916}{\texttt{arXiv:1503.05916}}.

\bibitem{Abdallah:2015ter}
\hrefCMSnoop {}{J.~Abdallah { et~al.}, ``Simplified models for dark matter
  searches at the {LHC}'',} \textit{ Phys. Dark Univ.} \textbf{ 9-10} (2015) 8,
  \href{http://dx.doi.org/10.1016/j.dark.2015.08.001}{\doi{10.1016/j.dark.2015.08.001}},
\href{http://www.arXiv.org/abs/1506.03116}{\texttt{arXiv:1506.03116}}.

\bibitem{Abercrombie:2015wmb}
\hrefCMSnoop {}{D.~Abercrombie { et~al.}, ``Dark matter benchmark models for
  early {LHC} {Run-2} searches: Report of the {ATLAS/CMS} dark matter forum'',}
  \textit{ Phys. Dark Univ.} \textbf{ 26} (2019) 100371,
  \href{http://dx.doi.org/10.1016/j.dark.2019.100371}{\doi{10.1016/j.dark.2019.100371}},
\href{http://www.arXiv.org/abs/1507.00966}{\texttt{arXiv:1507.00966}}.

\bibitem{BOVEIA2020100365}
\hrefCMSnoop {}{{A. Boveia et. al.}, ``{Recommendations on presenting {LHC}
  searches for missing transverse energy signals using simplified {$s$-channel}
  models of dark matter}'',} \textit{ Phys. Dark. Univ.} \textbf{ 27} (2020)
  100365,
  \href{http://dx.doi.org/10.1016/j.dark.2019.100365}{\doi{10.1016/j.dark.2019.100365}},
  \href{http://www.arXiv.org/abs/1603.04156}{\texttt{arXiv:1603.04156}}.

\bibitem{ATLAS2010}
\hrefCMSnoop {}{{ATLAS Collaboration}, ``Search for new particles in two-jet
  final states in 7 {TeV} proton-proton collisions with the {ATLAS} detector at
  the {LHC}'',} \textit{ Phys. Rev. Lett.} \textbf{ 105} (2010) 161801,
  \href{http://dx.doi.org/10.1103/PhysRevLett.105.161801}{\doi{10.1103/PhysRevLett.105.161801}},
\href{http://www.arXiv.org/abs/1008.2461}{\texttt{arXiv:1008.2461}}.

\bibitem{Khachatryan:2010jd}
\hrefCMSnoop {}{{CMS Collaboration}, ``Search for dijet resonances in 7 {TeV}
  pp collisions at {CMS}'',} \textit{ Phys. Rev. Lett.} \textbf{ 105} (2010)
  211801,
  \href{http://dx.doi.org/10.1103/PhysRevLett.105.211801}{\doi{10.1103/PhysRevLett.105.211801}},
  \href{http://www.arXiv.org/abs/1010.0203}{\texttt{arXiv:1010.0203}}.
[Erratum \DOI{10.1103/PhysRevLett.106.029902}].

\bibitem{Aad:2011aj}
\hrefCMSnoop {}{{ATLAS Collaboration}, ``{Search for new physics in dijet mass
  and angular distributions in pp collisions at $\sqrt{s} = 7$ TeV measured
  with the ATLAS detector}'',} \textit{ New J. Phys.} \textbf{ 13} (2011)
  053044,
  \href{http://dx.doi.org/10.1088/1367-2630/13/5/053044}{\doi{10.1088/1367-2630/13/5/053044}},
\href{http://www.arXiv.org/abs/1103.3864}{\texttt{arXiv:1103.3864}}.

\bibitem{Chatrchyan2011123}
\hrefCMSnoop {}{{CMS Collaboration}, ``{Search for resonances in the dijet mass
  spectrum from 7 TeV pp collisions at CMS}'',} \textit{ Phys. Lett. B}
  \textbf{ 704} (2011) 123,
  \href{http://dx.doi.org/10.1016/j.physletb.2011.09.015}{\doi{10.1016/j.physletb.2011.09.015}},
\href{http://www.arXiv.org/abs/1107.4771}{\texttt{arXiv:1107.4771}}.

\bibitem{Aad201237}
\hrefCMSnoop {}{{ATLAS Collaboration}, ``Search for new physics in the dijet
  mass distribution using 1 fb$^{-1}$ of pp collision data at $\sqrt{s}$ = 7
  {TeV} collected by the {ATLAS} detector'',} \textit{ Phys. Lett. B} \textbf{
  708} (2012) 37,
  \href{http://dx.doi.org/10.1016/j.physletb.2012.01.035}{\doi{10.1016/j.physletb.2012.01.035}},
\href{http://www.arXiv.org/abs/1108.6311}{\texttt{arXiv:1108.6311}}.

\bibitem{ATLAS:2012pu}
\hrefCMSnoop {}{{ATLAS Collaboration}, ``{ATLAS search for new phenomena in
  dijet mass and angular distributions using pp collisions at $\sqrt{s}=7$
  TeV}'',} \textit{ JHEP} \textbf{ 01} (2013) 029,
  \href{http://dx.doi.org/10.1007/JHEP01(2013)029}{\doi{10.1007/JHEP01(2013)029}},
  \href{http://www.arXiv.org/abs/1210.1718}{\texttt{arXiv:1210.1718}}.

\bibitem{CMS:2012yf}
\hrefCMSnoop {}{{CMS Collaboration}, ``{Search for narrow resonances and
  quantum black holes in inclusive and b-tagged dijet mass spectra from pp
  collisions at $\sqrt{s}=7$ TeV}'',} \textit{ JHEP} \textbf{ 01} (2013) 013,
  \href{http://dx.doi.org/10.1007/JHEP01(2013)013}{\doi{10.1007/JHEP01(2013)013}},
  \href{http://www.arXiv.org/abs/1210.2387}{\texttt{arXiv:1210.2387}}.

\bibitem{Chatrchyan:2013qhXX}
\hrefCMSnoop {}{{CMS Collaboration}, ``{Search for narrow resonances using the
  dijet mass spectrum in pp collisions at $\sqrt{s}$ = 8 TeV}'',} \textit{
  Phys. Rev. D} \textbf{ 87} (2013) 114015,
  \href{http://dx.doi.org/10.1103/PhysRevD.87.114015}{\doi{10.1103/PhysRevD.87.114015}},
  \href{http://www.arXiv.org/abs/1302.4794}{\texttt{arXiv:1302.4794}}.

\bibitem{Aad:2014aqa}
\hrefCMSnoop {}{{ATLAS Collaboration}, ``{Search for new phenomena in the dijet
  mass distribution using pp collision data at $\sqrt{s}=8$ TeV with the ATLAS
  detector}'',} \textit{ Phys. Rev. D} \textbf{ 91} (2015) 052007,
  \href{http://dx.doi.org/10.1103/PhysRevD.91.052007}{\doi{10.1103/PhysRevD.91.052007}},
\href{http://www.arXiv.org/abs/1407.1376}{\texttt{arXiv:1407.1376}}.

\bibitem{Khachatryan:2015sja}
\hrefCMSnoop {}{{CMS Collaboration}, ``{Search for resonances and quantum black
  holes using dijet mass spectra in proton-proton collisions at $\sqrt{s} = 8$
  TeV}'',} \textit{ Phys. Rev. D} \textbf{ 91} (2015) 052009,
  \href{http://dx.doi.org/10.1103/PhysRevD.91.052009}{\doi{10.1103/PhysRevD.91.052009}},
\href{http://www.arXiv.org/abs/1501.04198}{\texttt{arXiv:1501.04198}}.

\bibitem{ATLAS:2015nsi}
\hrefCMSnoop {}{{ATLAS Collaboration}, ``{Search for new phenomena in dijet
  mass and angular distributions from pp collisions at $\sqrt{s}=$ 13 TeV with
  the ATLAS detector}'',} \textit{ Phys. Lett. B} \textbf{ 754} (2016) 302,
  \href{http://dx.doi.org/10.1016/j.physletb.2016.01.032}{\doi{10.1016/j.physletb.2016.01.032}},
\href{http://www.arXiv.org/abs/1512.01530}{\texttt{arXiv:1512.01530}}.

\bibitem{Khachatryan:2015dcf}
\hrefCMSnoop {}{{CMS Collaboration}, ``Search for narrow resonances decaying to
  dijets in proton-proton collisions at {$\sqrt{s} =$ 13\TeV}'',} \textit{
  Phys. Rev. Lett.} \textbf{ 116} (2016) 071801,
  \href{http://dx.doi.org/10.1103/PhysRevLett.116.071801}{\doi{10.1103/PhysRevLett.116.071801}},
\href{http://www.arXiv.org/abs/1512.01224}{\texttt{arXiv:1512.01224}}.

\bibitem{Sirunyan:2016iap}
\hrefCMSnoop {}{{CMS Collaboration}, ``{Search for dijet resonances in
  proton-proton collisions at $\sqrt{s}$ = 13 TeV and constraints on dark
  matter and other models}'',} \textit{ Phys. Lett. B} \textbf{ 769} (2017)
  520,
  \href{http://dx.doi.org/10.1016/j.physletb.2017.02.012}{\doi{10.1016/j.physletb.2017.02.012}},
\href{http://www.arXiv.org/abs/1611.03568}{\texttt{arXiv:1611.03568}}.

\bibitem{Aaboud:2017yvp}
\hrefCMSnoop {}{{ATLAS Collaboration}, ``{Search for new phenomena in dijet
  events using 37 fb$^{-1}$ of pp collision data collected at $\sqrt{s}=13$ TeV
  with the ATLAS detector}'',} \textit{ Phys. Rev. D} \textbf{ 96} (2017)
  052004,
  \href{http://dx.doi.org/10.1103/PhysRevD.96.052004}{\doi{10.1103/PhysRevD.96.052004}},
\href{http://www.arXiv.org/abs/1703.09127}{\texttt{arXiv:1703.09127}}.

\bibitem{Sirunyan:2018xlo}
\hrefCMSnoop {}{{CMS Collaboration}, ``{Search for narrow and broad dijet
  resonances in proton-proton collisions at $\sqrt{s}=$ 13 TeV and constraints
  on dark matter mediators and other new particles}'',} \textit{ JHEP} \textbf{
  08} (2018) 130,
  \href{http://dx.doi.org/10.1007/JHEP08(2018)130}{\doi{10.1007/JHEP08(2018)130}},
\href{http://www.arXiv.org/abs/1806.00843}{\texttt{arXiv:1806.00843}}.

\bibitem{Aad:2019hjw}
\hrefCMSnoop {}{{ATLAS Collaboration}, ``{Search for new resonances in mass
  distributions of jet pairs using 139 fb$^{-1}$ of pp collisions at
  $\sqrt{s}=13$ TeV with the ATLAS detector}'',} \textit{ JHEP} \textbf{ 03}
  (2020) 145,
  \href{http://dx.doi.org/10.1007/JHEP03(2020)145}{\doi{10.1007/JHEP03(2020)145}},
  \href{http://www.arXiv.org/abs/1910.08447}{\texttt{arXiv:1910.08447}}.

\bibitem{Sirunyan:2019vgj}
\hrefCMSnoop {}{{CMS Collaboration}, ``{Search for high mass dijet resonances
  with a new background prediction method in proton-proton collisions at
  $\sqrt{s} =$ 13 TeV}'',} \textit{ JHEP} \textbf{ 05} (2020) 033,
  \href{http://dx.doi.org/10.1007/JHEP05(2020)033}{\doi{10.1007/JHEP05(2020)033}},
  \href{http://www.arXiv.org/abs/1911.03947}{\texttt{arXiv:1911.03947}}.

\bibitem{CMS:2019xai}
\hrefCMSnoop {}{{CMS Collaboration}, ``Search for low-mass quark-antiquark
  resonances produced in association with a photon at $\sqrt{s}=$ 13 {TeV}'',}
  \textit{ Phys. Rev. Lett.} \textbf{ 123} (2019) 231803,
  \href{http://dx.doi.org/10.1103/PhysRevLett.123.231803}{\doi{10.1103/PhysRevLett.123.231803}},
  \href{http://www.arXiv.org/abs/1905.10331}{\texttt{arXiv:1905.10331}}.

\bibitem{ATLAS:2018hbc}
\hrefCMSnoop {}{{ATLAS Collaboration}, ``{Search for light resonances decaying
  to boosted quark pairs and produced in association with a photon or a jet in
  proton-proton collisions at $\sqrt{s}=13$ TeV with the ATLAS detector}'',}
  \textit{ Phys. Lett. B} \textbf{ 788} (2019) 316,
  \href{http://dx.doi.org/10.1016/j.physletb.2018.09.062}{\doi{10.1016/j.physletb.2018.09.062}},
  \href{http://www.arXiv.org/abs/1801.08769}{\texttt{arXiv:1801.08769}}.

\bibitem{CMS:2019emo}
\hrefCMSnoop {}{{CMS Collaboration}, ``{Search for low mass vector resonances
  decaying into quark-antiquark pairs in proton-proton collisions at
  $\sqrt{s}=$ 13 TeV}'',} \textit{ Phys. Rev. D} \textbf{ 100} (2019) 112007,
  \href{http://dx.doi.org/10.1103/PhysRevD.100.112007}{\doi{10.1103/PhysRevD.100.112007}},
  \href{http://www.arXiv.org/abs/1909.04114}{\texttt{arXiv:1909.04114}}.

\bibitem{Khachatryan:2016ecr}
\hrefCMSnoop {}{{CMS Collaboration}, ``Search for narrow resonances in dijet
  final states at {$\sqrt{s}=$ 8\TeV} with the novel {CMS} technique of data
  scouting'',} \textit{ Phys. Rev. Lett.} \textbf{ 117} (2016) 031802,
  \href{http://dx.doi.org/10.1103/PhysRevLett.117.031802}{\doi{10.1103/PhysRevLett.117.031802}},
\href{http://www.arXiv.org/abs/1604.08907}{\texttt{arXiv:1604.08907}}.

\bibitem{ATLAS:2018qto}
\hrefCMSnoop {}{{ATLAS Collaboration}, ``{Search for low-mass dijet resonances
  using trigger-level jets with the ATLAS detector in pp collisions at
  $\sqrt{s}=13$ TeV}'',} \textit{ Phys. Rev. Lett.} \textbf{ 121} (2018)
  081801,
  \href{http://dx.doi.org/10.1103/PhysRevLett.121.081801}{\doi{10.1103/PhysRevLett.121.081801}},
  \href{http://www.arXiv.org/abs/1804.03496}{\texttt{arXiv:1804.03496}}.

\bibitem{2024scoutingRun3}
\hrefCMSnoop {}{{CMS Collaboration}, ``{Enriching the physics program of the
  CMS experiment via data scouting and data parking}'',} \textit{ Phys. Rept.}
  \textbf{ 1115} (2025) 678,
  \href{http://dx.doi.org/10.1016/j.physrep.2024.09.006}{\doi{10.1016/j.physrep.2024.09.006}},
  \href{http://www.arXiv.org/abs/2403.16134}{\texttt{arXiv:2403.16134}}.

\bibitem{HEPData}
\hrefCMSnoop {}{}{HEPD}ata record for this analysis, 2025.
\newblock
  \href{http://dx.doi.org/10.17182/hepdata.159438}{\doi{10.17182/hepdata.159438}}.

\bibitem{refCMS}
\hrefCMSnoop {}{{CMS Collaboration}, ``The {CMS} experiment at the {CERN}
  {LHC}'',} \textit{ JINST} \textbf{ 3} (2008) S08004,
\href{http://dx.doi.org/10.1088/1748-0221/3/08/S08004}{\doi{10.1088/1748-0221/3/08/S08004}}.

\bibitem{Hayrapetyan_2024_Run3}
\hrefCMSnoop {}{{CMS Collaboration}, ``Development of the {CMS} detector for
  the {CERN} {LHC} {Run 3}'',} \textit{ JINST} \textbf{ 19} (2024) P05064,
  \href{http://dx.doi.org/10.1088/1748-0221/19/05/p05064}{\doi{10.1088/1748-0221/19/05/p05064}},
  \href{http://www.arXiv.org/abs/2309.05466}{\texttt{arXiv:2309.05466}}.

\bibitem{Cacciari:2005hq}
\hrefCMSnoop {}{M.~Cacciari and G.~P. Salam, ``Dispelling the {$N^{3}$} myth
  for the {$k_\mathrm{T}$} jet-finder'',} \textit{ Phys. Lett. B} \textbf{ 641}
  (2006) 57,
  \href{http://dx.doi.org/10.1016/j.physletb.2006.08.037}{\doi{10.1016/j.physletb.2006.08.037}},
\href{http://www.arXiv.org/abs/hep-ph/0512210}{\texttt{arXiv:hep-ph/0512210}}.

\bibitem{Cacciari:2008gp}
\hrefCMSnoop {}{M.~Cacciari, G.~P. Salam, and G.~Soyez, ``{The
  anti-$k_\mathrm{T}$ jet clustering algorithm}'',} \textit{ JHEP} \textbf{ 04}
  (2008) 063,
  \href{http://dx.doi.org/10.1088/1126-6708/2008/04/063}{\doi{10.1088/1126-6708/2008/04/063}},
  \href{http://www.arXiv.org/abs/0802.1189}{\texttt{arXiv:0802.1189}}.

\bibitem{Cacciari:2011ma}
\hrefCMSnoop {}{M.~Cacciari, G.~P. Salam, and G.~Soyez, ``{FastJet user
  manual}'',} \textit{ Eur. Phys. J. C} \textbf{ 72} (2012) 1896,
  \href{http://dx.doi.org/10.1140/epjc/s10052-012-1896-2}{\doi{10.1140/epjc/s10052-012-1896-2}},
\href{http://www.arXiv.org/abs/1111.6097}{\texttt{arXiv:1111.6097}}.

\bibitem{Cacciari:2007fd}
\hrefCMSnoop {}{M.~Cacciari and G.~P. Salam, ``{Pileup subtraction using jet
  areas}'',} \textit{ Phys. Lett. B} \textbf{ 659} (2008) 119,
  \href{http://dx.doi.org/10.1016/j.physletb.2007.09.077}{\doi{10.1016/j.physletb.2007.09.077}},
  \href{http://www.arXiv.org/abs/0707.1378}{\texttt{arXiv:0707.1378}}.

\bibitem{Khachatryan:2016kdb}
\hrefCMSnoop {}{{CMS Collaboration}, ``{Jet energy scale and resolution in the
  CMS experiment in pp collisions at 8 TeV}'',} \textit{ JINST} \textbf{ 12}
  (2017) P02014,
  \href{http://dx.doi.org/10.1088/1748-0221/12/02/P02014}{\doi{10.1088/1748-0221/12/02/P02014}},
\href{http://www.arXiv.org/abs/1607.03663}{\texttt{arXiv:1607.03663}}.

\bibitem{Khachatryan:2016bia}
\hrefCMSnoop {}{{CMS Collaboration}, ``{The CMS trigger system}'',} \textit{
  JINST} \textbf{ 12} (2017) P01020,
  \href{http://dx.doi.org/10.1088/1748-0221/12/01/P01020}{\doi{10.1088/1748-0221/12/01/P01020}},
\href{http://www.arXiv.org/abs/1609.02366}{\texttt{arXiv:1609.02366}}.

\bibitem{Hayrapetyan_2024_Run2}
\hrefCMSnoop {}{{CMS Collaboration}, ``Performance of the {CMS} high-level
  trigger during {LHC Run 2}'',} \textit{ JINST} \textbf{ 19} (2024) P11021,
  \href{http://dx.doi.org/10.1088/1748-0221/19/11/p11021}{\doi{10.1088/1748-0221/19/11/p11021}},
  \href{http://www.arXiv.org/abs/2410.17038}{\texttt{arXiv:2410.17038}}.

\bibitem{CMS:2020ebo}
\hrefCMSnoop {}{{CMS Collaboration}, ``{Pileup mitigation at CMS in 13 TeV
  data}'',} \textit{ JINST} \textbf{ 15} (2020) P09018,
  \href{http://dx.doi.org/10.1088/1748-0221/15/09/P09018}{\doi{10.1088/1748-0221/15/09/P09018}},
  \href{http://www.arXiv.org/abs/2003.00503}{\texttt{arXiv:2003.00503}}.

\bibitem{CMS:2017wyc}
\href {https://cds.cern.ch/record/2256875}{{CMS Collaboration}, ``{Jet
  algorithms performance in 13 TeV data}'',} CMS Physics Analysis Summary
  CMS-PAS-JME-16-003, 2017.

\bibitem{Sjostrand:2014zea}
T.~Sj{\"o}strand\hrefCMSnoop {}{ { et~al.}, ``An introduction to {PYTHIA}
  8.2'',} \textit{ Comput. Phys. Commun.} \textbf{ 191} (2015) 159,
  \href{http://dx.doi.org/10.1016/j.cpc.2015.01.024}{\doi{10.1016/j.cpc.2015.01.024}},
\href{http://www.arXiv.org/abs/1410.3012}{\texttt{arXiv:1410.3012}}.

\bibitem{Skands:2014pea}
\hrefCMSnoop {}{P.~Skands, S.~Carrazza, and J.~Rojo, ``{Tuning PYTHIA 8.1: the
  Monash 2013 tune}'',} \textit{ Eur. Phys. J. C} \textbf{ 74} (2014) 3024,
  \href{http://dx.doi.org/10.1140/epjc/s10052-014-3024-y}{\doi{10.1140/epjc/s10052-014-3024-y}},
  \href{http://www.arXiv.org/abs/1404.5630}{\texttt{arXiv:1404.5630}}.

\bibitem{Khachatryan:2015pea}
\hrefCMSnoop {}{{CMS Collaboration}, ``{Event generator tunes obtained from
  underlying event and multiparton scattering measurements}'',} \textit{ Eur.
  Phys. J. C} \textbf{ 76} (2016) 155,
  \href{http://dx.doi.org/10.1140/epjc/s10052-016-3988-x}{\doi{10.1140/epjc/s10052-016-3988-x}},
\href{http://www.arXiv.org/abs/1512.00815}{\texttt{arXiv:1512.00815}}.

\bibitem{refGEANT}
\hrefCMSnoop {}{{GEANT4} Collaboration, ``{GEANT4} --- a simulation toolkit'',}
  \textit{ Nucl. Instrum. Meth. A} \textbf{ 506} (2003) 250,
\href{http://dx.doi.org/10.1016/S0168-9002(03)01368-8}{\doi{10.1016/S0168-9002(03)01368-8}}.

\bibitem{Harris:2011bh}
\hrefCMSnoop {}{R.~M. Harris and K.~Kousouris, ``{Searches for dijet resonances
  at hadron colliders}'',} \textit{ Int. J. Mod. Phys. A} \textbf{ 26} (2011)
  5005,
  \href{http://dx.doi.org/10.1142/S0217751X11054905}{\doi{10.1142/S0217751X11054905}},
\href{http://www.arXiv.org/abs/1110.5302}{\texttt{arXiv:1110.5302}}.

\bibitem{CMS:2022usq}
\hrefCMSnoop {}{{CMS Collaboration}, ``{Search for resonant and nonresonant
  production of pairs of dijet resonances in proton-proton collisions at
  $\sqrt{s} =$ 13 TeV}'',} \textit{ JHEP} \textbf{ 07} (2023) 161,
  \href{http://dx.doi.org/10.1007/JHEP07(2023)161}{\doi{10.1007/JHEP07(2023)161}},
  \href{http://www.arXiv.org/abs/2206.09997}{\texttt{arXiv:2206.09997}}.

\bibitem{ATLAS2025Dijet}
\hrefCMSnoop {}{{ATLAS Collaboration}, ``Search for electroweak-scale dijet
  resonances using trigger-level analysis with the {ATLAS} detector in 132
  fb$^{-1}$ of $pp$ collisions at $\sqrt{s}=13$ {TeV}'',} \textit{ Phys. Rev.
  D} \textbf{ 112} (2025) 092015,
  \href{http://dx.doi.org/10.1103/15p2-bkg8}{\doi{10.1103/15p2-bkg8}},
  \href{http://www.arXiv.org/abs/2509.01219}{\texttt{arXiv:2509.01219}}.

\bibitem{Fisher1922}
\hrefCMSnoop {}{{R. A. Fisher}, ``On the interpretation of $\chi^2$ from
  contingency tables, and the calculation of p'',} \textit{ J. Roy. Stat. Soc.}
  \textbf{ 85} (1922)
  \href{http://dx.doi.org/10.2307/2340521}{\doi{10.2307/2340521}}.

\bibitem{Sirunyan:2759951}
\hrefCMSnoop {}{{CMS Collaboration}, ``{Precision luminosity measurement in
  proton-proton collisions at $\sqrt{s} =$ 13 TeV in 2015 and 2016 at CMS}'',}
  \textit{ Eur. Phys. J. C} \textbf{ 81} (2021) 800,
  \href{http://dx.doi.org/10.1140/epjc/s10052-021-09538-2}{\doi{10.1140/epjc/s10052-021-09538-2}},
  \href{http://www.arXiv.org/abs/2104.01927}{\texttt{arXiv:2104.01927}}.

\bibitem{CMS-PAS-LUM-17-004}
\href {https://cds.cern.ch/record/2621960}{{CMS Collaboration}, ``{CMS}
  luminosity measurements for the 2017 data taking period'',} CMS Physics
  Analysis Summary CMS-PAS-LUM-17-004, 2017.

\bibitem{CMS-PAS-LUM-18-002}
\href {https://cds.cern.ch/record/2676164}{{CMS Collaboration}, ``{CMS}
  luminosity measurements for the 2018 data taking period'',} CMS Physics
  Analysis Summary CMS-PAS-LUM-18-002, 2017.

\bibitem{Junk:1999kv}
\hrefCMSnoop {}{T.~Junk, ``{Confidence level computation for combining searches
  with small statistics}'',} \textit{ Nucl. Instrum. Meth. A} \textbf{ 434}
  (1999) 435,
  \href{http://dx.doi.org/10.1016/S0168-9002(99)00498-2}{\doi{10.1016/S0168-9002(99)00498-2}},
  \href{http://www.arXiv.org/abs/hep-ex/9902006}{\texttt{arXiv:hep-ex/9902006}}.

\bibitem{Read:2002hq}
\hrefCMSnoop {}{A.~L. Read, ``{Presentation of search results: the
  {CL$\mathrm{_s}$} technique}'',} \textit{ {J. Phys. G}} \textbf{ {28}}
  ({2002}) {2693},
  \href{http://dx.doi.org/10.1088/0954-3899/28/10/313}{\doi{10.1088/0954-3899/28/10/313}}.

\bibitem{CMS:2024onh}
\hrefCMSnoop {}{{CMS Collaboration}, ``The {CMS} statistical analysis and
  combination tool: {Combine}'',} \textit{ Comput. Softw. Big Sci.} \textbf{ 8}
  (2024) 19,
  \href{http://dx.doi.org/10.1007/s41781-024-00121-4}{\doi{10.1007/s41781-024-00121-4}},
  \href{http://www.arXiv.org/abs/2404.06614}{\texttt{arXiv:2404.06614}}.

\bibitem{Verkerke:2003ir}
\href
  {https://www.slac.stanford.edu/econf/C0303241/proc/papers/MOLT007.PDF}{W.~Verkerke
  and D.~Kirkby, ``The \textsc{RooFit} toolkit for data modeling'',} in
  \textit{ {\it Proc. 13th Int. Conf. on Computing in High Energy and Nuclear
  Physics (CHEP): La Jolla CA, United States}}.
\newblock 2003.
\newblock
  \href{http://www.arXiv.org/abs/physics/0306116}{\texttt{arXiv:physics/0306116}}.

\bibitem{Moneta:2010pm}
L.~Moneta\hrefCMSnoop {}{ { et~al.}, ``The \textsc{RooStats} project'',} in
  \textit{ {\it Proc. 13th Int. Workshop on Advanced Computing and Analysis
  Techniques in Physics Research (ACAT): Jaipur, India}}.
\newblock 2010.
\newblock \href{http://www.arXiv.org/abs/1009.1003}{\texttt{arXiv:1009.1003}}.
\newblock
  \href{http://dx.doi.org/10.22323/1.093.0057}{\doi{10.22323/1.093.0057}}.

\bibitem{Cowan:2010js}
\hrefCMSnoop {}{G.~Cowan, K.~Cranmer, E.~Gross, and O.~Vitells, ``Asymptotic
  formulae for likelihood-based tests of new physics'',} \textit{ Eur. Phys. J.
  C} \textbf{ 71} (2011) 1554,
  \href{http://dx.doi.org/10.1140/epjc/s10052-011-1554-0}{\doi{10.1140/epjc/s10052-011-1554-0}},
  \href{http://www.arXiv.org/abs/1007.1727}{\texttt{arXiv:1007.1727}}.
[Erratum: \DOI{10.1140/epjc/s10052-013-2501-z}].

\end{thebibliography}\endgroup

\cleardoublepage \appendix\section{The CMS Collaboration \label{app:collab}}\begin{sloppypar}\hyphenpenalty=5000\widowpenalty=500\clubpenalty=5000
\cmsinstitute{Yerevan Physics Institute, Yerevan, Armenia}
{\tolerance=6000
A.~Hayrapetyan, V.~Makarenko\cmsorcid{0000-0002-8406-8605}, A.~Tumasyan\cmsAuthorMark{1}\cmsorcid{0009-0000-0684-6742}
\par}
\cmsinstitute{Institut f\"{u}r Hochenergiephysik, Vienna, Austria}
{\tolerance=6000
W.~Adam\cmsorcid{0000-0001-9099-4341}, J.W.~Andrejkovic, L.~Benato\cmsorcid{0000-0001-5135-7489}, T.~Bergauer\cmsorcid{0000-0002-5786-0293}, M.~Dragicevic\cmsorcid{0000-0003-1967-6783}, C.~Giordano, P.S.~Hussain\cmsorcid{0000-0002-4825-5278}, M.~Jeitler\cmsAuthorMark{2}\cmsorcid{0000-0002-5141-9560}, N.~Krammer\cmsorcid{0000-0002-0548-0985}, A.~Li\cmsorcid{0000-0002-4547-116X}, D.~Liko\cmsorcid{0000-0002-3380-473X}, M.~Matthewman, I.~Mikulec\cmsorcid{0000-0003-0385-2746}, J.~Schieck\cmsAuthorMark{2}\cmsorcid{0000-0002-1058-8093}, R.~Sch\"{o}fbeck\cmsAuthorMark{2}\cmsorcid{0000-0002-2332-8784}, D.~Schwarz\cmsorcid{0000-0002-3821-7331}, M.~Shooshtari\cmsorcid{0009-0004-8882-4887}, M.~Sonawane\cmsorcid{0000-0003-0510-7010}, W.~Waltenberger\cmsorcid{0000-0002-6215-7228}, C.-E.~Wulz\cmsAuthorMark{2}\cmsorcid{0000-0001-9226-5812}
% ...
\par}
\vskip\cmsinstskip
\dag:~Deceased\\
$^{1}$Also at Yerevan State University, Yerevan, Armenia\\
$^{2}$Also at TU Wien, Vienna, Austria\\
% ...
\end{sloppypar}
\end{document}